\documentclass{article}
\pdfoutput=1
\usepackage{amsmath}
\usepackage{amssymb}
\usepackage{graphicx}
\usepackage{color}
\usepackage[usenames,dvipsnames,svgnames,table]{xcolor}
\usepackage{soul}
\usepackage{subfig}
\usepackage{ulem}

\pdfpageheight\paperheight
\pdfpagewidth\paperwidth

\usepackage{jcappub}

\definecolor{darkred}{RGB}{175,0,0}

\newcommand{\eeq}{\end{equation}} 
\newcommand{\Msun}{M_\odot}
\newcommand{\fpbh}{f_{\rm PBH}}
\usepackage{array}
\def\ga{ \mathrel{\rlap{\raise 0.59ex
 \hbox{$>$}}{\lower 0.59ex \hbox{$\sim$}}}}
\begin{document}
\title{Cosmological implications of Primordial Black Holes}
\author[a,b]{Jos\' e Luis Bernal}
\author[a,b]{Nicola Bellomo}
\author[a]{Alvise Raccanelli}
\author[a,c]{Licia Verde}

\affiliation[a]{ICC, University of Barcelona, IEEC-UB, Mart\' i i Franqu\` es, 1, E08028
Barcelona, Spain}
\affiliation[b]{Dept. de F\' isica Qu\` antica i Astrof\' isica, Universitat de Barcelona, Mart\' i i Franqu\` es 1, E08028 Barcelona,
Spain}
\affiliation[c]{ICREA, Pg. Llu\' is Companys 23, 08010 Barcelona, Spain}

\emailAdd{joseluis.bernal@icc.ub.edu}
\emailAdd{nicola.bellomo@icc.ub.edu}
\emailAdd{alvise@icc.ub.edu}
\emailAdd{liciaverde@icc.ub.edu}

\abstract{
The possibility that a relevant fraction of the dark matter might be comprised of Primordial Black Holes (PBHs) has been seriously reconsidered after LIGO's detection of a $\sim 30 M_{\odot}$ binary black holes merger. Despite the strong interest in the model, there is a lack of studies on 
possible cosmological implications and effects on cosmological parameters inference.
We investigate correlations with the other standard cosmological parameters using cosmic microwave background observations, finding significant degeneracies, especially with the tilt of the primordial power spectrum and the sound horizon at radiation drag. However, these degeneracies can be greatly reduced with the inclusion of  small scale polarization data.
We also explore if PBHs as dark matter in simple extensions of the standard $\Lambda$CDM cosmological model induces extra degeneracies, especially between the additional parameters and the PBH's ones.
Finally, we present  cosmic microwave background constraints on the fraction of dark matter in PBHs, not only for monochromatic PBH mass distributions but also for popular extended mass distributions. Our results show that extended mass distribution's constraints are tighter, but also that a considerable amount of constraining power comes from the high-$\ell$ polarization data.
Moreover, we constrain the shape of such mass distributions in terms of the correspondent constraints on the PBH mass fraction.
}
\maketitle

\hypersetup{pageanchor=true}

\section{Introduction}\label{sec:Introduction}
The concept of Primordial Black Holes (PBHs) was introduced in the sixties \cite{Zeldovich_pbh}, and subsequently it was suggested that they might make up the dark matter~\cite{Chapline_pbh}. However, increasingly stringent constraints (see e.g. \cite{Alcock_micro1998,Flynn_barydm97,Carr_dynamic1999,Wilkinson_smbhconstraints}) gave way to theories proposing elementary particles as dark matter. Of the latter, the most popular theory is the Weakly Interactive Massive Particles (see e.g., \cite{Jungman_wimps}). Nonetheless, the fact that WIMPs are still undetected while experiments are reaching the background sensitivity \cite{Arcadi_wimps} joint with the LIGO+VIRGO collaboration's \cite{abbott:ligo} first detection of gravitational waves emissions from $\sim 30 M_\odot$ binary black hole merger, make it timely to reconsider PBHs abundance constraints (see e.g.,~\cite{Bird_pbh, Sasaki_pbh, Clesse_cluspbh}).

It is important to bear in mind that even without accounting for a relevant fraction of the dark matter, the existence of PBHs might be a possible solution for other astrophysical open questions. For instance, PBHs might be the progenitors of the super massive black holes located at the nuclei of galaxies (e.g.,~\cite{Kohri_smbh,Bernal_SMBH} and references therein) or the intermediate massive black holes that could inhabit the center of dwarf galaxies (e.g.,~\cite{silk_dwarf} and references therein).

The abundance of PBHs (and hence the fraction of the total dark matter that they constitute, $f_{\rm PBH}$) is constrained by several independent observations in a wide range of masses, with present datasets and new observables suggested for future experiments (see e.g. \cite{raccanelli:cross, raccanelli:radio, raccanelli:quantumgravityconstraint, cholis, kovetz, munoz, carr:comparison2, Kuhnel_EMF,Schutz_pta,AliHaimoud_merger,Kovetz_gwpbh}).
For stellar masses, existing constraints include microlensing by compact objects with masses $\lesssim 10 \Msun$ \cite{griest:keplerconstraint,niikura:microlensingconstraint,tisserand:microlensingconstraint,calchinovati:microlensingconstraint,mediavilla:microlensingconstraint}, wide binaries disruption\cite{quinn:widebinaryconstraint} or stellar distribution in ultra-faint dwarfs galaxies \cite{brandt:ufdgconstraint} at slightly larger masses. There are also constraints on $\fpbh$ on this mass range from X-ray and radio observations of the Milky Way \cite{gaggero}, although they depend strongly on astrophysical assumptions.
In addition to astrophysical observables, the presence of PBHs has also consequences on cosmological observables such as the Cosmic Microwave Background (CMB).

The basic mechanism behind the effects that a PBH population has on the CMB is the following. PBHs accrete primordial gas in the early Universe and inject energy into the primordial plasma via radiation. Therefore, the Universe's thermal and ionization histories are affected by the presence of PBHs, leaving potentially detectable signatures in the CMB. Given that the medium has more energy because of the PBH energy injection, recombination is delayed. This shifts the acoustic peaks and affects some physical quantities (e.g., the sound horizon at radiation drag, $r_{\rm s}$). To summarize, the imprints are similar to those imposed by the energy injection of exotic species such as dark matter decaying into photons \cite{giesen:annihilatingdm}.

Recently, Ali-Ha\" imoud \& Kamionkowski \citep{Ali-Haimoud_PBH} (hereafter AHK) rederived constraints from CMB power spectra and spectral distortions using the recently released Planck power spectra \cite{Planckparameterspaper}, assuming spherical accretion and considering two limiting cases of ionization mechanisms (collisional ionization and photoionization). Compared to the previous analysis, AHK generalize the radiative efficiency computation accounting for Compton drag and cooling by CMB photons, as well as ionization cooling once the gas is neutral, and use a more precise estimate of the relative velocity between PBHs and baryons. All this leads to significantly smaller accretion rates and PBH luminosities than previous analyses, therefore weakening constraints with respect to previous studies \cite{ricotti:pbhcosmologicaleffect}. These updated results leave a window open for PBHs to be the dark matter ($f_{\rm PBH}=1$), precisely for masses of tens of $\Msun$. In contrast to the standard cosmological model, $\Lambda$-Cold Dark Matter ($\Lambda$CDM), hereafter we refer to a $\Lambda$CDM model where a significant fraction of the dark matter is PBH as $\Lambda$PBH. 
Recently, the authors of \cite{poulin:cmbconstraint} revisited CMB constraints assuming disk accretion. They find tighter constraints, which close the mentioned window and exclude the possibility that PBHs of tens of $\Msun$ account for $f_{\rm PBH}\ga 0.1$. This would rule out the  $\Lambda$PBH model.  However, both spherical accretion and disk accretion for all the PBHs are limiting cases. It is therefore reasonable to expect that a realistic scenario would be in between these (Refs.\cite{Ali-Haimoud_PBH} and \cite{poulin:cmbconstraint}) limiting cases. 

Most of the constraints mentioned above are derived assuming a Dirac delta function for the mass distribution (i.e., {\it monochromatic} distribution). While being a good approximation as a first step, this is an idealized case. Extended mass distributions (EMDs)
 appear naturally in formation mechanisms such as the collapse of large primordial fluctuations \cite{carr:pbhfrominhomogeneities} or cosmic strings \cite{hawking:pbhformation}, or cosmological phase transitions, like bubble collisions \cite{hawking:pbhfrombubbles}, among others \cite{Clesse_pbhhybrid}. Besides, critical collapse \cite{Choptuik_criticcollapse} broadens any distribution, even if it is nearly monochromatic, making an EMD for PBHs unavoidable \cite{Niemeyer_criticcollapse}. This effect has been studied numerically in \cite{Musco_numcc05,Musco_numcc09,Musco_numcc13}, showing that it applies over ten orders of magnitude in density contrast (see e.g. figure 1 of \cite{Musco_numcc09}).  
 
 Given the wide variety of EMDs for PBHs, it is impossible to explore them all in detail in a cosmological context.
 The authors of \cite{carr:comparison1,carr:comparison2,Kuhnel_EMF} have proposed approaches to translate monochromatic constraints to EMDs; for a discussion of the advantages and limitations of these approaches see the above references and \cite{bellomo:pbhemfconstraints}. Constraints for simple EMDs have been derived exactly for microlensing observations \cite{green:lognormal}.

Recent works demonstrate that it is possible to interpret very accurately the effects of a population of PBHs with an EMD in each different cosmological probe as a monochromatic population with a corresponding effective equivalent mass. The approach is physically motivated as it accounts for the underlying physics of the effects of PBHs \cite{bellomo:pbhemfconstraints}. This is the approach we follow here. We refer the reader to Ref. \cite{bellomo:pbhemfconstraints} for more details and advantages compared to other approaches.

Although recently there has been a renewed effort to revisit PBHs constraints, there is still a lack of studies on PBHs cosmological implications and their possible correlations with other cosmological parameters. Given the persistent tensions that exist among some observations at high and low redshift within $\Lambda$CDM, it is worthwhile to explore the possibility that the inclusion of PBHs in the model might reconcile these tensions. 
Using the formalism described in AHK, we study the impact of a large PBH mass fraction on CMB-derived cosmological parameters, for the standard six parameters of $\Lambda$CDM and for common extensions to this model. We also compute constraints on $\fpbh$ for PBHs with EMDs and test the prescriptions of \cite{bellomo:pbhemfconstraints} for the effective equivalent mass for the CMB.

This paper is structured as follows. In Section \ref{sec:MethandData} we introduce the observational data we use and our methodology. Results are presented in Section \ref{sec:Results}: the results for monochromatic distributions are shown in Section \ref{sec:Results_monochromatic} and those for EMDs, in Section \ref{sec:EMF}. Finally, discussion and conclusions can be found in Section \ref{sec:Conclusions}.

\section{Methodology and Data}\label{sec:MethandData}
We use the public Boltzmann code CLASS \cite{Lesgourgues:2011re,Blas:2011rf} using the modified version of HyRec \cite{alihamoud:hyrec1, alihamoud:hyrec2} introduced in AHK. This modification allows us to compute CMB power spectra accounting for effects due to a monochromatic mass distribution of PBH with mass $M$ and fraction $f_\mathrm{PBH}$. We further modify this code to allow also for a variety of PBH EMDs following the prescriptions suggested in AHK. 

We use the Markov Chain Monte Carlo (MCMC) public code Monte Python \cite{Audren13_mp} to infer cosmological parameters using the observational data described below. We use uniform priors for $\fpbh$ in the range $0\leq\fpbh\leq 1$, which is motivated by the linear dependence of the differences in the CMB power spectra on $\fpbh$, shown in AHK. We also consider logarithmic priors finding that results do no change significantly.

We consider the full Planck 2015 temperature (TT), polarization (EE) and the cross correlation of temperature and polarization (TE) angular power spectra \cite{Planckparameterspaper}, corresponding to the following likelihoods: Planck high$\ell$ TTTEEE (for $\ell\geq 30$), Planck low$\ell$ (for $2\leq \ell \leq 29$) and the lensing power spectrum (CMB lensing). The Planck team identifies the low$\ell$ + high$\ell$ TT as the recommended baseline dataset for models beyond $\Lambda$CDM and the high$\ell$ polarization data as preliminary, because of evidence of low level systematics ($\sim {\rm \mu K^2}$ in $\ell(\ell+1)C_\ell$) \cite{Planck15_likelihood}.
While the level of systematic contamination does not appear to affect parameter estimation, we present results both excluding and including the high$\ell$ polarization data. Hereinafter we refer to the data set of Planck high$\ell$ TTTEEE, Planck low$\ell$ and CMB lensing as ``Full Planck" (or {\it full} for short) and we refer to the Planck recommended baseline as ``PlanckTT+lowP+lensing" (or {\it baseline}).

The approach we follow builds on AHK work and is similar in spirit.  
In that work, the authors use the \texttt{Plik\_lite} best fit $C_\ell$, a provided covariance matrix for the high$\ell$ CMB-only TTTEEE power spectra, and a prior on the optical depth of reionization, $\tau_{\rm reio}$, from \cite{Planck_newHFI}. Note that such covariance matrix is computed for a $\Lambda$CDM model without the presence of PBHs, so no correlation among $\fpbh$ and the standard cosmological parameters is considered.
One difference comes from the fact that, while AHK use a Fisher matrix approach to estimate parameters constraints, we use MCMC for parameter inference. The price of being more rigorous comes at a considerable increase in the computation time of the analysis. Computing the constraints on $\fpbh$ for all masses would take an unpractical amount of time. However, given that the dependence of the upper limits on $\fpbh$ on $M_{\rm PBH}$ is smooth, we choose $M_{\rm PBH}$ values equally spaced in logarithmic scale and interpolate the constraints in between.
With this procedure, for any $M_{\rm PBH}$ in the interval covered by the sampling, we have an estimated value for the 95\% and 68\% confidence limit on $\fpbh$. This correspondence will be useful in Sec.~\ref{sec:EMF} and Figs.~\ref{fig:2D_LN} and ~\ref{fig:2D_PL}.

\subsection{Accounting for EMDs}\label{sec:EMDmeth}
To consider the effect of an extended mass distribution (EMD), rather than exploring the whole parameter space of all possibilities, we follow the Bellomo et al. \cite{bellomo:pbhemfconstraints} prescriptions to interpret EMDs as monochromatic populations with an effective equivalent mass, $M_{\rm eq}$, which depends on both the shape of the EMD and the observable considered.  
Here we just briefly introduce the formalism for the CMB and refer the interested reader to Ref.~\cite{bellomo:pbhemfconstraints} for a full explanation. 

For an EMD, we define the PBH mass fraction as:
\begin{equation}
\frac{d \fpbh}{dM} = \fpbh \frac{d\Phi_{\rm{PBH}}}{dM},
\label{eq:dfdM}
\end{equation}
in such a way that $\frac{d\Phi_{\rm{PBH}}}{dM}$ is normalized to unity. We consider two popular mass distributions: a power law (PL) and a lognormal (LN). The former can be expressed as:
\begin{equation}
\frac{d\Phi_\mathrm{PBH}}{dM} = \frac{\mathcal{N}_{\rm PL}(\gamma, M_\mathrm{min}, M_\mathrm{max})}{M^{1-\gamma}}\Theta(M-M_\mathrm{min})\Theta(M_\mathrm{max}-M),
\end{equation}
characterized by the exponent $\gamma$, a mass range $(M_\mathrm{min}, M_\mathrm{max})$ and a normalization factor
\begin{equation}
\mathcal{N}_{\rm PL}(\gamma, M_\mathrm{min}, M_\mathrm{max}) = \left\lbrace
\begin{aligned}
&\frac{\gamma}{M^\gamma_\mathrm{max}-M^\gamma_\mathrm{min}}, &\gamma\neq0,	\\
&\log^{-1}\left(\frac{M_\mathrm{max}}{M_\mathrm{min}}\right),	&\gamma=0,
\end{aligned}\right.
\label{eq:normPL}
\end{equation}
where $\gamma=-\frac{2w}{1+w}$, $w$ being the equation of state when PBHs form. If an expanding Universe is assumed ($w>-1/3$), $\gamma$ spans the range $(-1,1)$.
 We consider two cases of power law distribution: one with $\gamma = 0$ (corresponding to PBHs formed in matter domination epoch) and another one with $\gamma = -0.5$ (corresponding to PBHs formed in radiation dominated epoch).

Lognormal mass distributions can be expressed as:
\begin{equation}
\frac{d\Phi_{\rm{PBH}}}{dM} = \frac{1}{\sqrt{2\pi\sigma^2M^2}}e^{-\frac{\log^2(M/\mu)}{2\sigma^2}}\,,
\end{equation}
where $\mu$ and $\sigma$ are the mean and the standard deviation of the logarithm of the mass.

In order to account for an EMD, the energy density injection rate has to be integrated over the whole mass range spanned by PBHs:
\begin{equation}
\dot{\rho}_\mathrm{inj}=\rho_\mathrm{dm}f_\mathrm{PBH}\int dM\frac{d\Phi_{\rm PBH}}{dM}\frac{\left\langle L(M)\right\rangle}{M},
\label{eq:cmb_total_injected_energy}
\end{equation}
where $\left\langle L(M)\right\rangle$ is the velocity-averaged\footnote{Recall that the accretion rate depends, among others, on the relative velocity PBH-gas.} luminosity of a PBH. Starting from the results of AHK, it is possible to estimate the mass dependence of the integrand in Eq. \eqref{eq:cmb_total_injected_energy} as:
\begin{equation}
\frac{\left\langle L\right\rangle}{M} \propto \frac{L}{M} \propto \frac{\dot{M}^2/L_\mathrm{Edd}}{M} \propto \frac{M^4\lambda^2(M)/M}{M} = M^2\lambda^2(M),
\end{equation}
where $L$ is the luminosity of an accreting black hole, $\dot{M}$ is the black hole growth rate, $L_\mathrm{Edd}$ is the Eddington luminosity and $\lambda(M)$ is the dimensionless accretion rate. In principle the averaged luminosity will depend not only on the mass but also on redshift, gas temperature, free electron fraction
and ionization regime. Ref.~\cite{bellomo:pbhemfconstraints} assumes that these dependencies can be factored out by parametrizing the dimensionless accretion rate as $\lambda(M)=M^{\alpha/2}$, where $\alpha$ is a parameter tuned numerically a posteriori to minimize differences in the relevant
observable quantity between the EMD case and the equivalent monochromatic case.

Therefore, the effects of an EMD on the CMB power spectra are mimicked by a monochromatic distribution of mass $M_{\rm eq}$ when: 
\begin{equation}
f^\mathrm{MMD}_\mathrm{PBH}M_\mathrm{eq}^{2+\alpha} = f^\mathrm{EMD}_\mathrm{PBH}\left\lbrace
\begin{aligned}
&\mathcal{N}_{\rm PL}\frac{M^{\gamma+2+\alpha}_\mathrm{max}-M^{\gamma+2+\alpha}_\mathrm{min}}{\gamma+2+\alpha},	&PL,\\
&\mu^{2+\alpha} e^{\frac{(2+\alpha)^2\sigma^2}{2}},	&LN.
\end{aligned}\right.
\label{eq:cmb_equivalence}
\end{equation}
where $\fpbh^{\rm MMD}$ is the PBH mass fraction associated with a monochromatic distribution. In order to obtain the effective equivalent mass, we fix $\fpbh^{\rm MMD} = \fpbh^{\rm EMD}$. This way, the upper limit on $\fpbh^{\rm EMD}$ is the corresponding upper limit for a monochromatic distribution with the effective equivalent mass.

As the authors of \cite{bellomo:pbhemfconstraints} indicate, there is no single choice of the parameter $\alpha$ in Eq.~\ref{eq:cmb_equivalence} able to correctly match the time dependent energy injection of the EMDs with the monochromatic case. Since we require to select one effective value for $\alpha$, the relation between the ``best" value of $\alpha$ and the redshift range where the match is optimised depends also on the EMD. Given that Planck power spectra have the smallest error bars for $10^2 \lesssim \ell \lesssim 10^3$, we choose to select $\alpha$ by minimizing the differences in the CMB power spectra in that multipole range. Although we expect that different values of $\alpha$ do not have great impact in the performance of the method, we use $\alpha = 0.3$ for Power Law distributions, and $\alpha = 0.2$ for Lognormal distributions \cite{bellomo:pbhemfconstraints}. 

In principle, our effective treatment of the EMD (Eq.~\ref{eq:cmb_equivalence}) and related results strictly applies only for $M< 10^4\Msun$. 
This is because AHK modelling includes a steady-state approximation to avoid time dependent fluid, heat and ionization computations. This approximation holds when the characteristic accretion timescale is much shorter than the Hubble timescale, which is fulfilled for PBHs with $M\lesssim 3\times 10^4\Msun$ \cite{Ricotti_accretion}. Since AHK formalism breaks down for $M>10^4\Msun$, Eq.~\eqref{eq:cmb_equivalence} is valid only for EMDs which do not have significant contributions beyond that limit.
 
 \subsection{Important considerations}\label{sec:caveats}
Several considerations are in order before we present our results and attempt their interpretation.
First of all, the treatment of the effects of PBHs on the CMB (adopted from AHK) only considers changes in the ionization and Hydrogen thermal history via energy injection due to PBHs accretion. It does not consider any changes in the background history of the Universe. These changes would be mainly introduced because of the energy transfer from the matter component to the radiation one through black hole mergers, for instance.  However there are several indications that these  processes are not important  in this context. In fact Ref. \cite{marti:gwpbhmergers} estimated that  no more than 1$\%$ of the dark matter can be converted into gravitational waves after recombination;
 hence this mechanism cannot provide a significant ``dark radiation" component affecting the early time expansion history and cosmological parameters estimation. It cannot be invoked therefore to reduce the tension between the inferred value of $H_0$ obtained using CMB observations and assuming $\Lambda$CDM \cite{Planckparameterspaper} and its direct measurement coming from the distance ladder \cite{RiessH0_2016}. 
However, black hole mergers could affect the expansion history before recombination in a similar way than $N_{\rm eff}$. This effect would further affect recombination quantities, such as $r_{\rm s}$, which also affects the inferred value of $H_0$ \cite{BernalH0, StandardQuantities}. This avenue is left to study in future work.

Energy transfer from matter to radiation sectors is not the only effect of PBH mergers. When two PBHs merge, the final state is a single black hole with larger mass. Thus the merger history of the PBH population changes the initial EMD predicted by inflation theories. The extent of this effect depends on the merger rate which in turn depends on  the initial clustering and mass and velocity distributions. Given current theoretical uncertainties and observational precision, neglecting changes in the EMD through time is not expected to bias current constraints. However, this may need to be accounted for in the future.

 As PBHs behave like a cold dark matter fluid at sufficiently large scales, they only affect cosmological observables via energy injection because of the accreted matter. Therefore, the only other cosmological probe affected by PBHs is reionization. The rest of cosmological probes (i.e., cosmic shear, baryon acoustic oscillations, etc) will not be affected by a large $\fpbh$. For this reason we have not considered other cosmological probes here. In practical applications they could certainly be used to further reduce parameter degeneracies.

While we assume spherical accretion, in \cite{poulin:cmbconstraint} disk accretion is assumed. As the radiative efficiency when a disk is formed is much larger than if the accretion is spherical, their constraints are much stronger than ours and those appearing in AHK. However, both scenarios are possible in the early Universe: whether the accretion is spherical or a disk is formed depends on the value of the angular momentum of the accreted gas at the Bondi radius compared with the angular momentum at the innermost stable circular orbit. Given that it is difficult to estimate the angular momentum of the gas accreting onto a PBH, the best approach is to consider these as two limiting cases. We refer the reader to Figure 3 and related discussion in Poulin et al. (2017) \cite{poulin:cmbconstraint} in order to see the extent of the differences in the shape of the power spectra between both scenarios. 
Nonetheless, we expect that
our results can be extrapolated to the disk accretion framework.  Actually, the degeneracies between $\fpbh$ and other
parameters are expected to be approximately the same in both cases. 

Therefore, it should be clear from the above discussion that uncertainties in the modelling of processes such as accretion mechanisms of PBHs or velocity distributions imply that any constraint should be interpreted as an order of magnitude estimate, rather than a precise quantity.
However, the behaviour of parameters degeneracies as a function of the different data sets considered or ionization regime should be qualitatively captured by our analysis. For this reason,  
we will also report the ratio between the constraints using two different data sets or ionization models.

As the equivalence relation between the effects of PBH with an EMD and a monochromatic distribution derived in \cite{bellomo:pbhemfconstraints} is obtained assuming the formalism of AHK, it is subject to the same caveats as AHK modelling. In particular given that AHK modelling breaks down for $M> 10^4 \Msun$, Eq.~\eqref{eq:cmb_equivalence} should be used strictly only for EMDs which do not have significant contributions beyond that limit. 
A criterion to decide if a EMD fullfil this condition can be found in Ref. \cite{bellomo:pbhemfconstraints}. 
Finally, while in principle the choice of the parameter $\alpha$ in Eq.~\ref{eq:cmb_equivalence} could be numerically optimized, the equivalent mass is relatively insensitive to small changes of its value for the EMDs considered here; therefore a slightly sub-optimal value of $\alpha$ does not bias our results.  

\section{Results}\label{sec:Results}
\subsection{Constraints on $\fpbh$ for monochromatic distributions}\label{sec:Results_monochromatic}
We start by directly comparing our approach with that of AHK; we use the same data (Planck lite high$\ell$TTTEEE + prior in $\tau$) but a MCMC instead of a Fisher approach to estimate parameter constraints.
 The constraints we obtain are similar to those of AHK, with variations always below the $10\%$ level, which is smaller than the theoretical uncertainty of the model. Although there is no significant effect introduced by using a Fisher approach, as we are interested in the posterior distribution in the whole parameter space, we use MCMC. 

\begin{figure}[ht]
\centering
\includegraphics[width=0.4\textwidth]{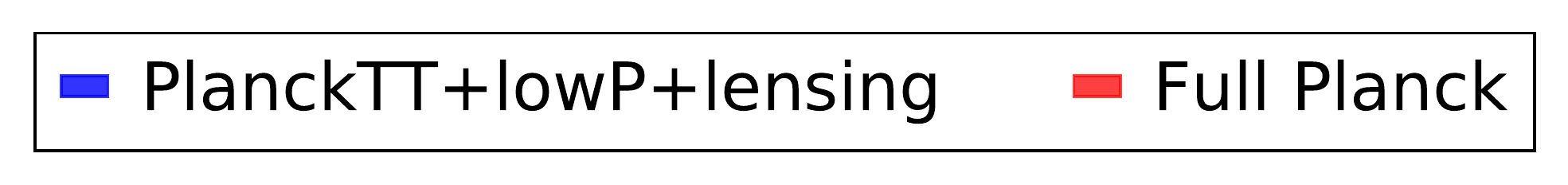} \\ 
\vspace{-0.1cm}
\includegraphics[width=0.9\textwidth]{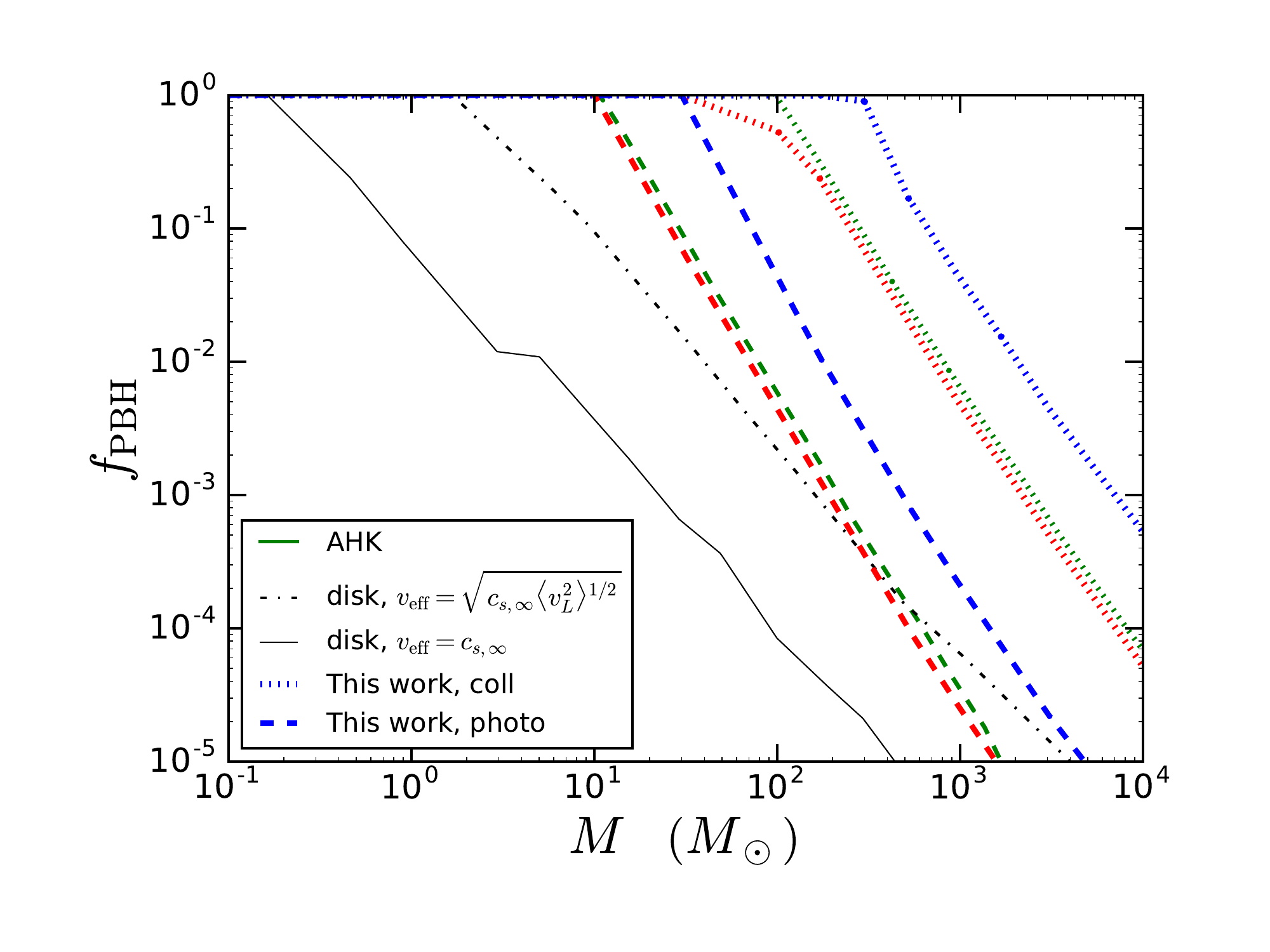}
\caption{68$\%$ confidence level marginalized constraints on $\fpbh$ in the collisional ionization regime (dotted lines) and in the photoionization regime (dashed lines). The results using the full data set of Planck are shown in red and without including the high multipoles of polarization, in blue. We also show, as a reference, the 68$\%$ confidence level marginalized constraints of AHK (green) and the 95$\%$ confidence level marginalized constraints of \cite{poulin:cmbconstraint} (black), where disk accretion is assumed. These external results are obtained including the high multipoles of Planck polarization power spectrum.
}
\label{fig:fpbh_constraints}
\end{figure}

In Figure \ref{fig:fpbh_constraints} we show the $68\%$ confidence level upper limits on $\fpbh$ for monochromatic populations of PBHs with different masses for both the full Planck dataset (red) and removing the high multipoles of polarization power spectrum (blue). Note that this color coding is used in all the figures throughout the paper. We also show the results of AHK and the 95$\%$ confidence level upper limits obtained assuming disk accretion from \cite{poulin:cmbconstraint}. The results using {\it full} Planck are very similar to those obtained in AHK. 
For the recommended baseline of Planck (which is the most conservative case) the constraints are weaker, widening the window where $\fpbh \sim 1$  is allowed. 
The maximum masses for which $\fpbh\sim 1$ is allowed (i.e., there is no upper limit at $68\%$ confidence level) in each of the cases considered are reported in Tab.~\ref{tab:fpbh_mono}.
\begin{table}[h!]
\small
\begin{center}
\begin{tabular}{|c|c|c|c|c|}
\hline
	& coll. Planck full 	& coll. baseline	& photo. Planck full & photo. baseline \\
	\hline
max. $M$ with $\fpbh\sim 1$	& 30$\Msun$ & 300$\Msun$ & 10$\Msun$ & 30$\Msun$ \\
\hline
\end{tabular}
\end{center}
\caption{
Maximum mass for which $\fpbh \sim 1$ is allowed at $68\%$ confidence level for monochromatic distributions. We report the values for all possible combinations of data sets (either full Planck or baseline) and ionization regimes (either collisional ionization or photoionization). Note that  (as it should be clear from the text and from Fig.~\ref{fig:fpbh_constraints}) these values are approximated and limited by the discrete sampling in mass for which we compute the constraints. 
}
\label{tab:fpbh_mono}
\end{table}

To capture the effect of different data sets and assumptions on the ionization regime on the $\fpbh$ limits, we use the ratio between two upper limits obtained with different assumptions or data.
Tab.~\ref{tab:rel_const_lcdm} highlights the effect of polarization high multipoles (first two columns) by reporting the ratio of the marginalised 68$\%$ confidence level on $\fpbh$ obtained using the {\it full} Planck data to that obtained using the {\it baseline} Planck data set. The next two columns illustrate the effect of the ionization regime (ratio of assuming photoionization over collisional ionization).
 As it can be seen, ratios for the different data sets does not depend on the ionization limit and ratios for the ionization limits does not depend on the data. This suggests that the ratio between two marginalized upper limits on $\fpbh$ can be used to isolate the effects of specific choices of modelling or data sets. 
In fact, including the high multipoles of polarization power spectra improves constraints by a factor of 10. On the other hand, assuming photoionization regime instead of collisional ionization, improve constraints by a factor of $\sim$200.

\begin{table}[h]
\small
\begin{center}
\begin{tabular}{|c|c|c|c|c|}
\hline
 & Planck full/ baseline & Planck full/baseline & 	photo/coll &	photo/coll \\
 & (coll)&(photo)& (Full Planck) & (baseline) \\
\hline
 $\fpbh$ ratio & 0.11	& 0.11	& 5.2$\times 10^{-3}$ & 4.8 $\times 10^{-3}$ \\ 
\hline
Validity 	& $M> 300\Msun$	& $M> 30\Msun$ & 	$M> 100\Msun$	& $M> 300\Msun$ \\
\hline
\end{tabular}
\end{center}
\caption{
Ratio of 68$\%$ marginalized upper limits on $\fpbh$. We compare the upper limits obtained using full Planck against using the baseline PlanckTT+lowP+lensing for the two ionization regimes and the constraints obtained assuming photoionization against assuming collisional ionization for both data sets. In all cases we report the mass range for which the correspondent ratio is valid.
}
\label{tab:rel_const_lcdm}
\end{table}

\subsubsection{Constraints and degeneracies with cosmological parameters in $\Lambda$CDM}
In addition to the constraints on $\fpbh$, we also study the degeneracies of this parameter with the six standard cosmological parameters of the $\Lambda$CDM model.

The most notable degeneracy, as already noticed in \cite{ricotti:pbhcosmologicaleffect}, is between $\fpbh$ and the scalar spectral index $n_s$. This positive correlation is induced by the energy injection of the PBH emission, which alters the tails of the CMB power spectra. As decoupling is delayed, the diffusion damping increases and the high multipoles are suppressed through Silk damping. 
\begin{figure}[t]
\centering
\includegraphics[width=0.4\textwidth]{Figures/legend_planck.pdf} \\ 
\includegraphics[width=\linewidth]{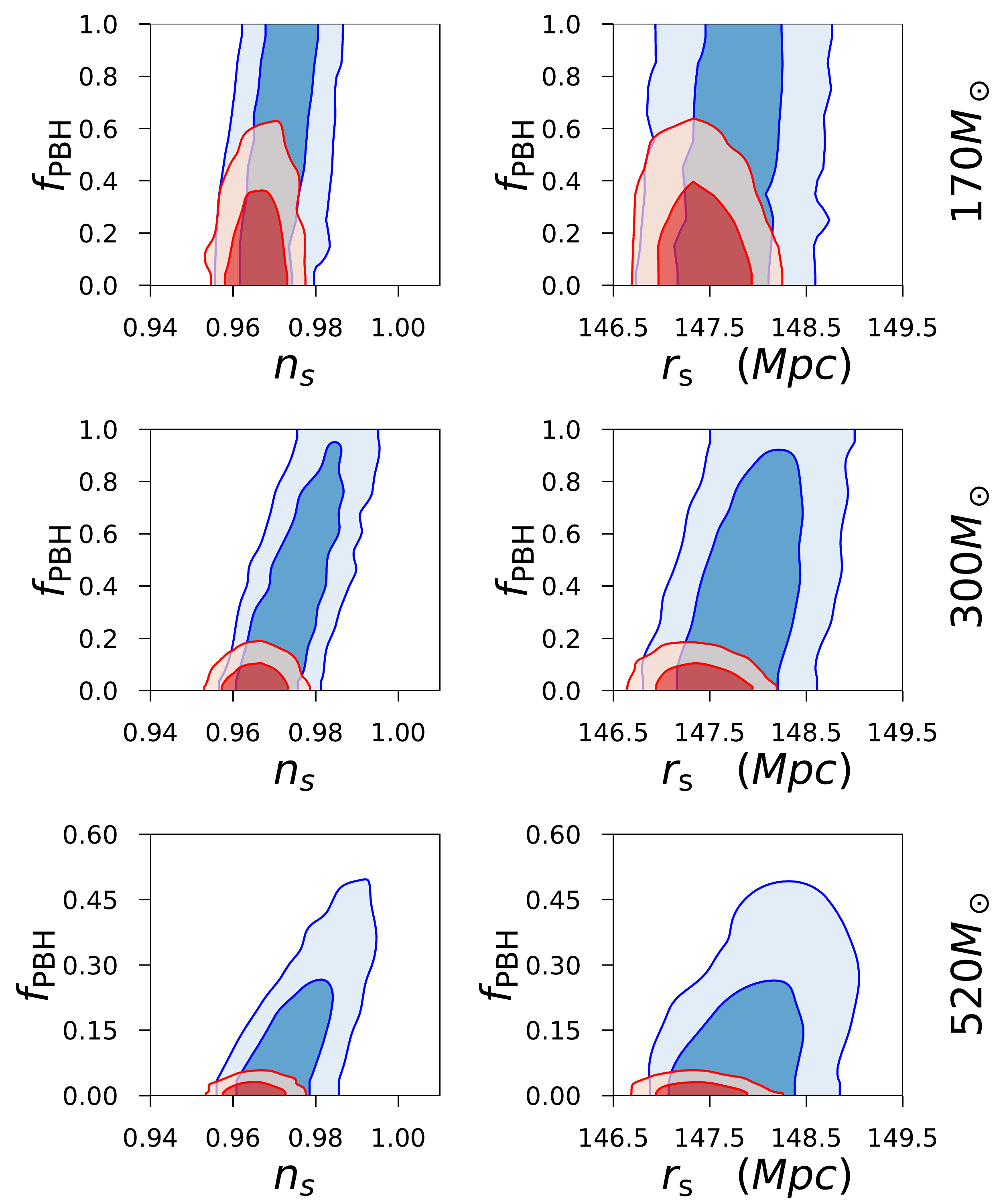}
\caption{68$\%$ and 95$\%$ confidence level marginalized constraints on the $n_s$-$\fpbh$ plane (left) and $r_{\rm s}$-$\fpbh$ plane (right)  for  $\Lambda$CDM with  PBHs as part of the dark matter (i.e., free $\fpbh$) in the collisional ionization regime. We consider three monochromatic distributions: 170$M_\odot$ (top), 300$M_\odot$ (middle) and 520$M_\odot$ (bottom). Results using the full data set of Planck are shown in red and using the recommended baseline, in blue. 68\% upper limits on $\fpbh$ correspond to the those shown in Figure \ref{fig:fpbh_constraints}. Note the change of scale in y-axis.}
\label{fig:coll_degen}
\end{figure}
\begin{figure}[t]
\centering
\includegraphics[width=0.4\textwidth]{Figures/legend_planck.pdf} \\ 
\includegraphics[width=\linewidth]{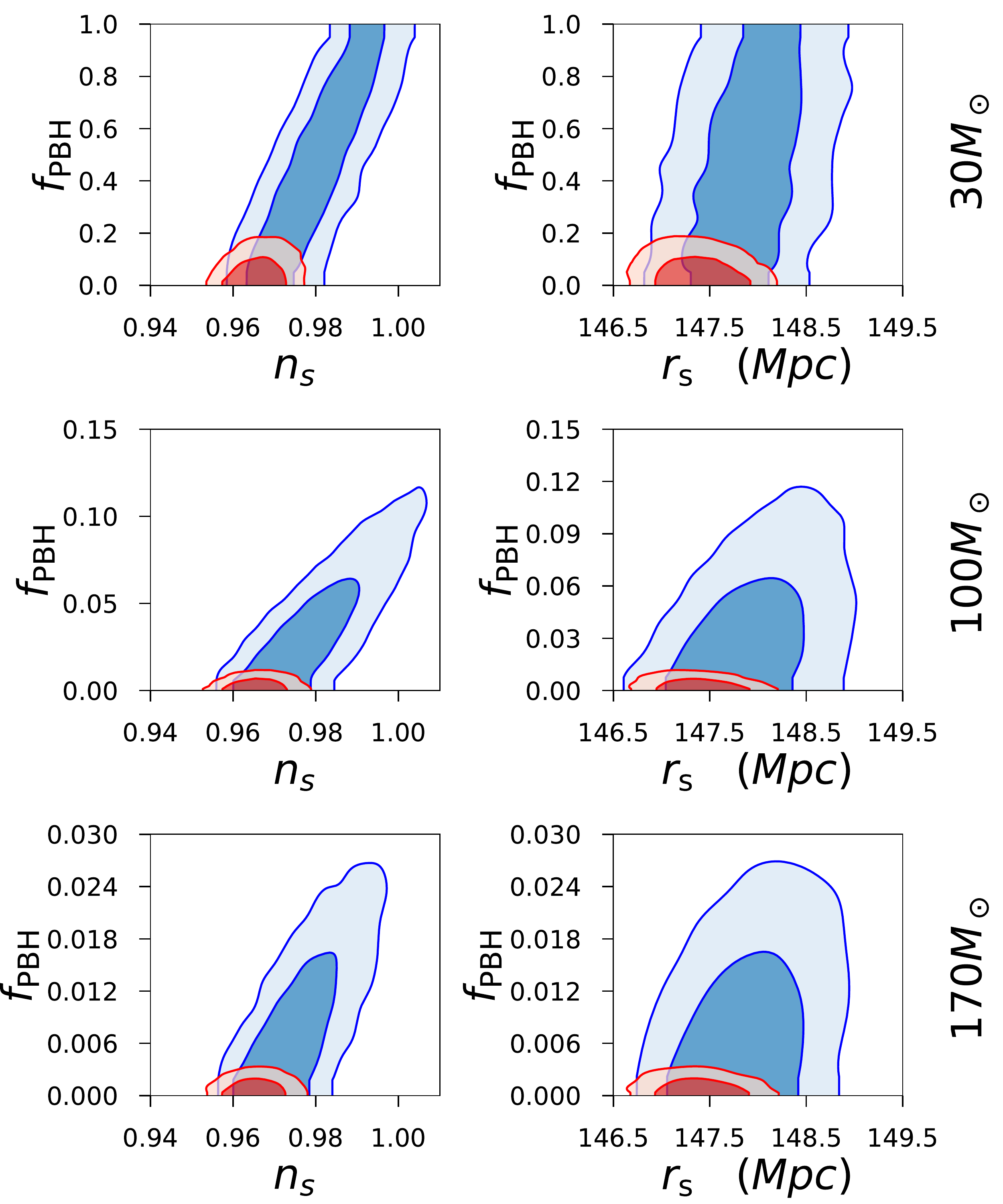}
\caption{Same as Fig.\ref{fig:coll_degen}, but in photoionization regime. In this case we consider 30$M_\odot$ (top), 100$M_\odot$ (middle) and 170$M_\odot$ (bottom).}
\label{fig:photo_degen}
\end{figure}
In Figures \ref{fig:coll_degen} and \ref{fig:photo_degen}, we show the results for the collisional ionization and photoionization limits, respectively. We report 68$\%$ and 95$\%$ confidence level marginalized constraints on the $\fpbh-n_s$ (left) and $\fpbh-r_{\rm s}$ (right) planes, for different masses, as these are the most affected parameters; in general, degeneracies become stronger for larger masses. 
For some cases, a scale independent or even blue tilted primordial power spectrum is allowed.
Combined analyses with galaxy surveys can be used to constrain $n_s$, reducing the degeneracy; we will discuss this in more details elsewhere.

The $\fpbh-r_{\rm s}$ degeneracy is the result of two competing effects. On one hand, the presence of PBHs delays recombination and shifts the peaks of CMB power spectra (see Figure 13 of AHK). This yields a larger sound horizon at radiation drag, leading to a lower value of $H_0$ to fit the cosmic distance ladder of supernovae type Ia and Baryon Acoustic Oscillations (BAO), as shown in \citep{BernalH0}. On the other hand, the effect of a delayed recombination on the value of $r_{\rm s}$ is compensated by small degeneracies in the density parameters of baryons and dark matter. Thus, a sizeable PBH component (with masses $\gtrsim 300\Msun$ for collisional ionization or $\gtrsim 30\Msun$ for photoionization) 
 can change $r_{\rm s}$ by $\sim 1$ Mpc. This shift is of a magnitude comparable to other effects and, most importantly, non-negligible with current and future errors in distance measurements using the BAO scale.

The effect on the rest of $\Lambda$CDM parameters is smaller, mostly resulting in mild relaxations of the confidence regions. Beyond that, only the amplitude of the scalar modes in the primordial power spectrum, $A_S$, presents a small positive correlation with $\fpbh$.\footnote{While in principle one may expect also a degeneracy with the optical depth to recombination $\tau_\mathrm{reio}$, this is not noticeable due to the tight constraints on this parameter imposed by the low $\ell$ polarization data and because $\fpbh$ is positive definite.}

Perhaps unsurprisingly, the inclusion of the high$\ell$ polarization data effectively remove the degeneracies; the effect on the TT damping tail can be mimicked by changes of the cosmological parameters, but this is not anymore the case once also the high$\ell$ polarization is considered.
It is well known that the high$\ell$ polarization data has a  similar effect in other models where, as in this model, the CMB damping tail is affected by the additional physics introduced (e.g., if dark radiation is allowed and for decaying dark matter models).

\subsubsection{Implications for extended cosmologies}\label{sec:Results_ext}
Here we explore the consequences of varying $\fpbh$ in some common extensions of the $\Lambda$CDM model. In addition to an expected weakening of the constraints, we study the degeneracies with the additional parameters. We consider the following models: free equation of state of dark energy ($w$CDM), free sum of neutrino masses ($\Lambda$CDM$+m_\nu$), free effective number of relativistic species ($\Lambda$CDM+$N_{\rm eff}$) and free running of the spectral index ($\Lambda$CDM$+\alpha_s$). Finally, we also allow the running of the running of the spectral index, $\beta_s$ to be a free parameter ($\Lambda$CDM$+\alpha_s$+$\beta_s$). We limit the study to the recommended baseline of Planck. Moreover, following the results from the previous section and in order to appreciate changes in $\fpbh$, we choose masses for which $\fpbh \sim 0.1$: 520$M_\odot$ for collisional ionization and 100$M_\odot$ for photoionization.
\begin{figure*}[htb!]
\centering
\includegraphics[width=0.8\linewidth]{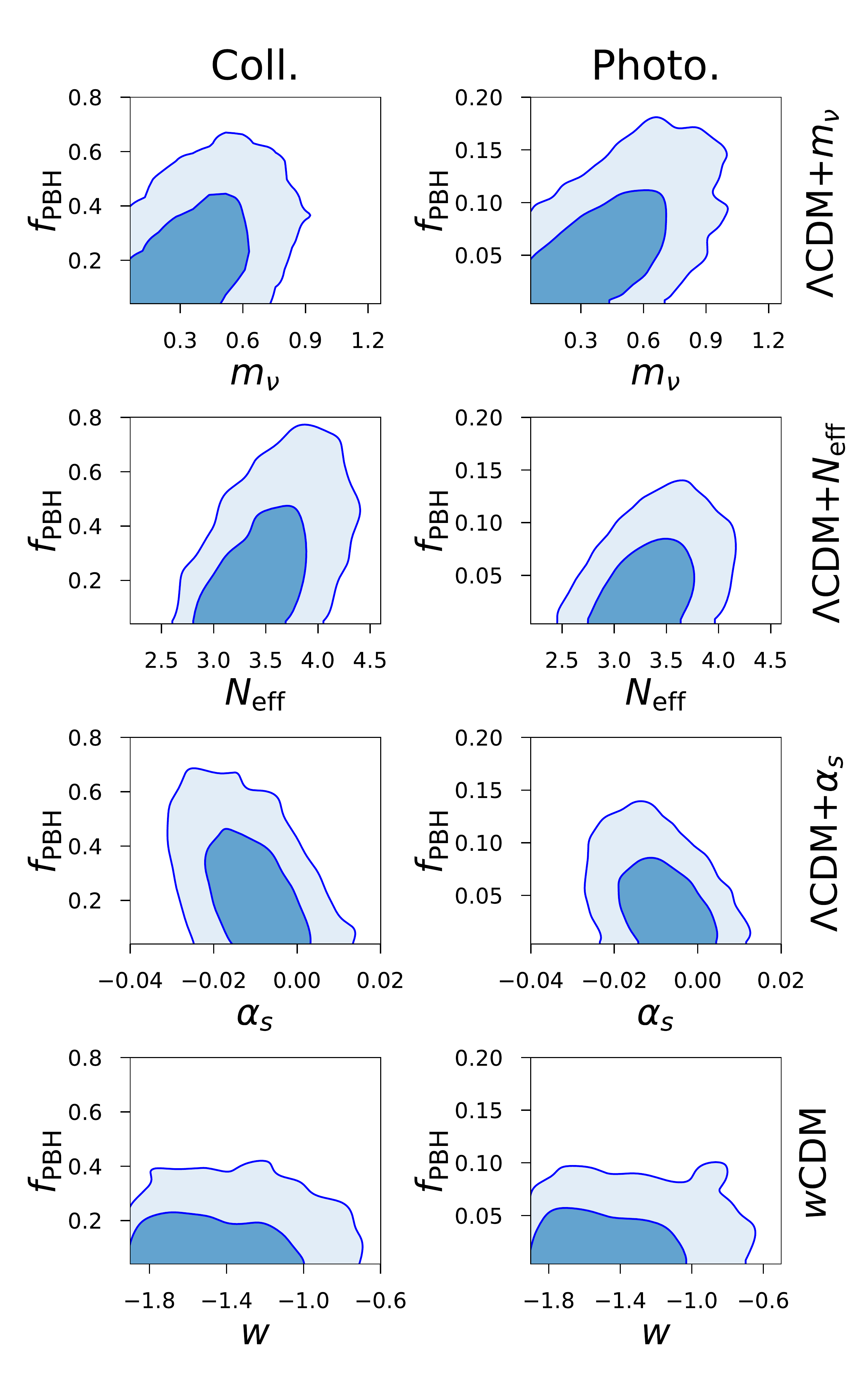}
\caption{68$\%$ and 95$\%$ confidence level marginalized constraints on the plane $\fpbh$-$x$ ($x$ being the additional parameter to $\Lambda$PBH) for 520$M_\odot$ in the collisional ionization regime (left) and 100$M_\odot$ in the photoionization (right) for the following extended cosmologies (form top to bottom): $\Lambda$PBH$+m_\nu$, $\Lambda$PBH+$N_{\rm eff}$, $\Lambda$PBH$+\alpha_s$ and $w$PBH.
}
\label{fig:extcosmo}
\end{figure*}

\begin{figure*}[htb!]
\centering
\includegraphics[width=\linewidth]{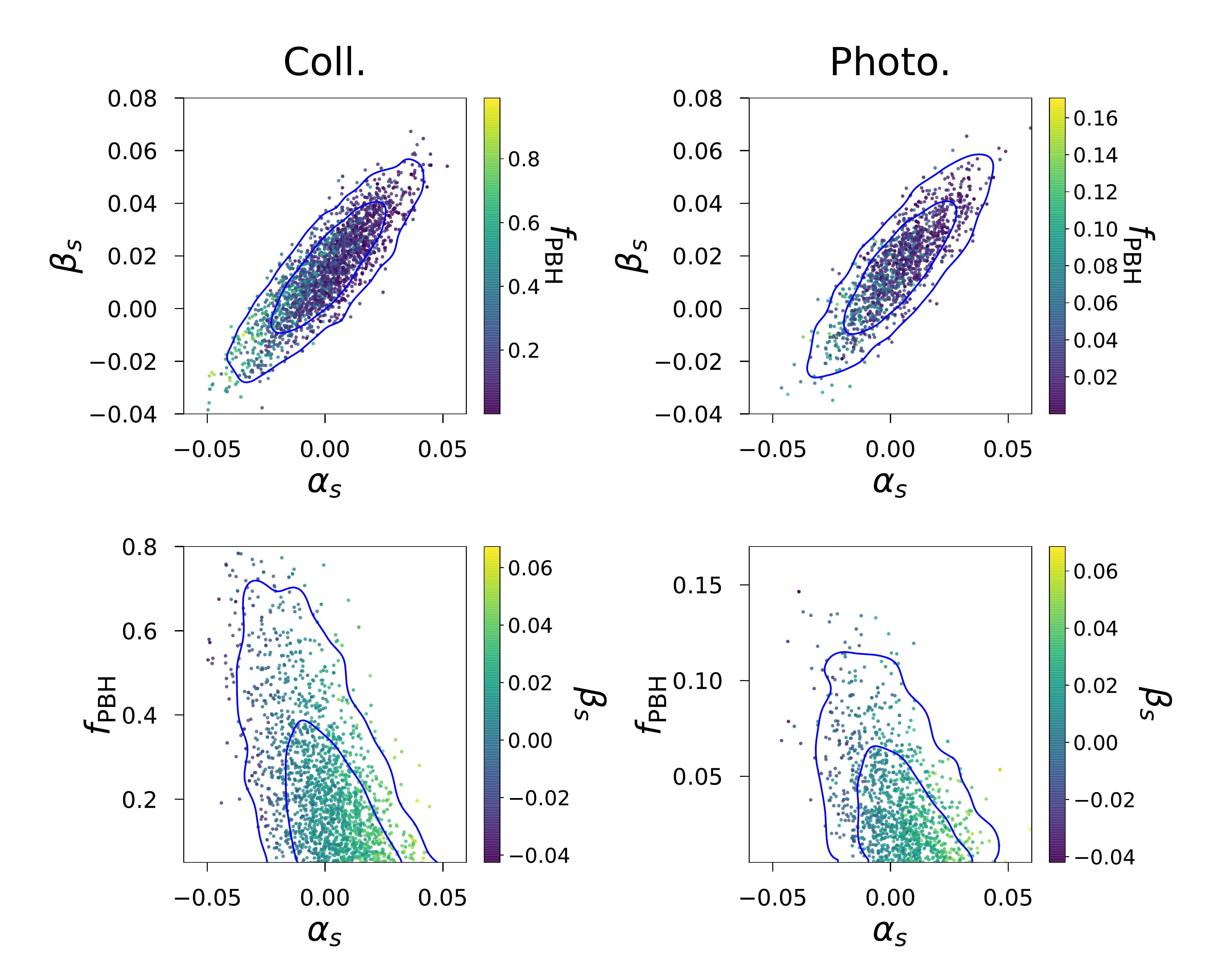}
\caption{68$\%$ and 95$\%$ confidence level marginalized constraints for a $\Lambda\mathrm{CDM}+\alpha_s+\beta_s$ model for 520$\Msun$ in the collisional regime (left panels) and for 100$\Msun$ in the photoionization regime (right panels). Upper panels: constraints on the  $\beta_s$-$\alpha_s$ plane with a color code to express the value of $\fpbh$. Bottom panels: constraints on the  $\fpbh$-$\alpha_s$ plane with a color code to indicate the value of $\beta_s$. These results are obtained without including the high multipoles of the polarization power spectrum of Planck. }
\label{fig:alphabeta}
\end{figure*}

Figure~\ref{fig:extcosmo} shows 68$\%$ and 95$\%$ confidence level marginalized constraints on the plane of $\fpbh$ and the additional parameters of the extended cosmologies. 
As it can be seen, there is a correlation between $m_{\nu}$ and $\fpbh$ and between $N_{\rm eff}$ and $\fpbh$, and an anti-correlation between $\alpha_s$ and $\fpbh$, but no appreciable correlation with $w$. The degeneracies with $m_\nu$, $N_{\rm eff}$ and $\alpha_s$ open the possibility that when limiting the theoretical framework to $\Lambda$CDM, one might hide the presence of PBHs. 
On the other hand, $\fpbh = 0$ is always within the $68\%$ confidence region. 

The positive correlation between $\fpbh$ and $N_{\rm eff}$ can be explained as follows. On one hand, as PBH energy injection delays recombination, $r_{\rm s}$ is higher (and CMB peaks are displaced towards larger scales). On the other hand, a value of $N_{\rm eff}$ different than the fiducial value changes the expansion history, especially in the early Universe, hence $r_{\rm s}$ is modified. Concretely, higher values of $N_{\rm eff}$ derive on lower values of $r_{\rm s}$ (and displacements of the CMB peaks towards smaller scales). This is why, if PBHs exist, larger values of $\fpbh$ need larger values of $N_{\rm eff}$ to fit observations. Varying $m_\nu$ also changes the expansion history, the angular diameter distance to recombination and the redshift of equality, so the relation between $\fpbh$ and $m_\nu$ is similar to that explained above.

We also consider the case in which the running of the spectral index, $\alpha_s$, is scale dependent, hence there is an additional free parameter in this case: the running of the running, $\beta_s$, which we consider scale independent. As it can be seen in Figure \ref{fig:alphabeta}, varying $\beta_s$ weakens the constraints on $\alpha_s$ and the parameters are highly correlated. What is more surprising is that for small values of $\fpbh$, $\beta_s$ is slightly different than 0. However, this trend is not strong enough to be statistically significant.  

Although there is a strong correlation between $\fpbh$ and $n_s$, the correlations with $\alpha_s$ and $\beta_s$ are much smaller. This is due to the suppression of the CMB power spectra on the high $\ell$ tail produced by PBHs can be mimicked by a different constant value of $n_s$, without evidence of the need of a strong scale dependent variation.

\begin{figure}[ht!]
\centering
\includegraphics[width=0.7\textwidth]{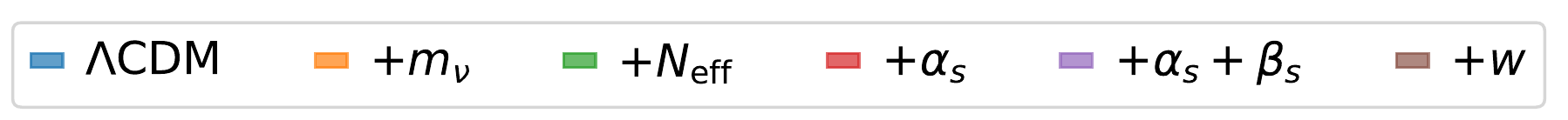}

\includegraphics[width=0.45\linewidth]{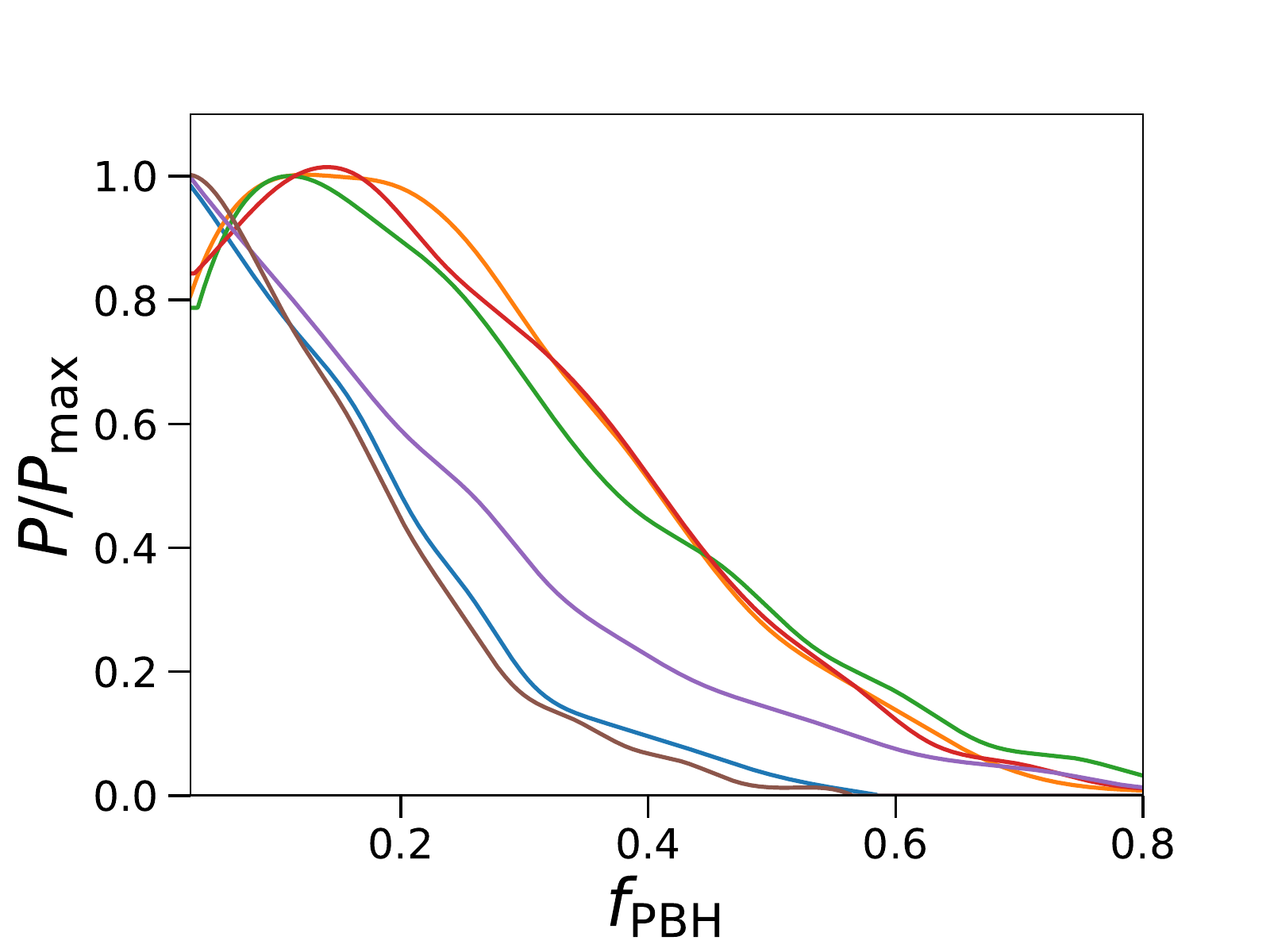}
\includegraphics[width=0.45\linewidth]{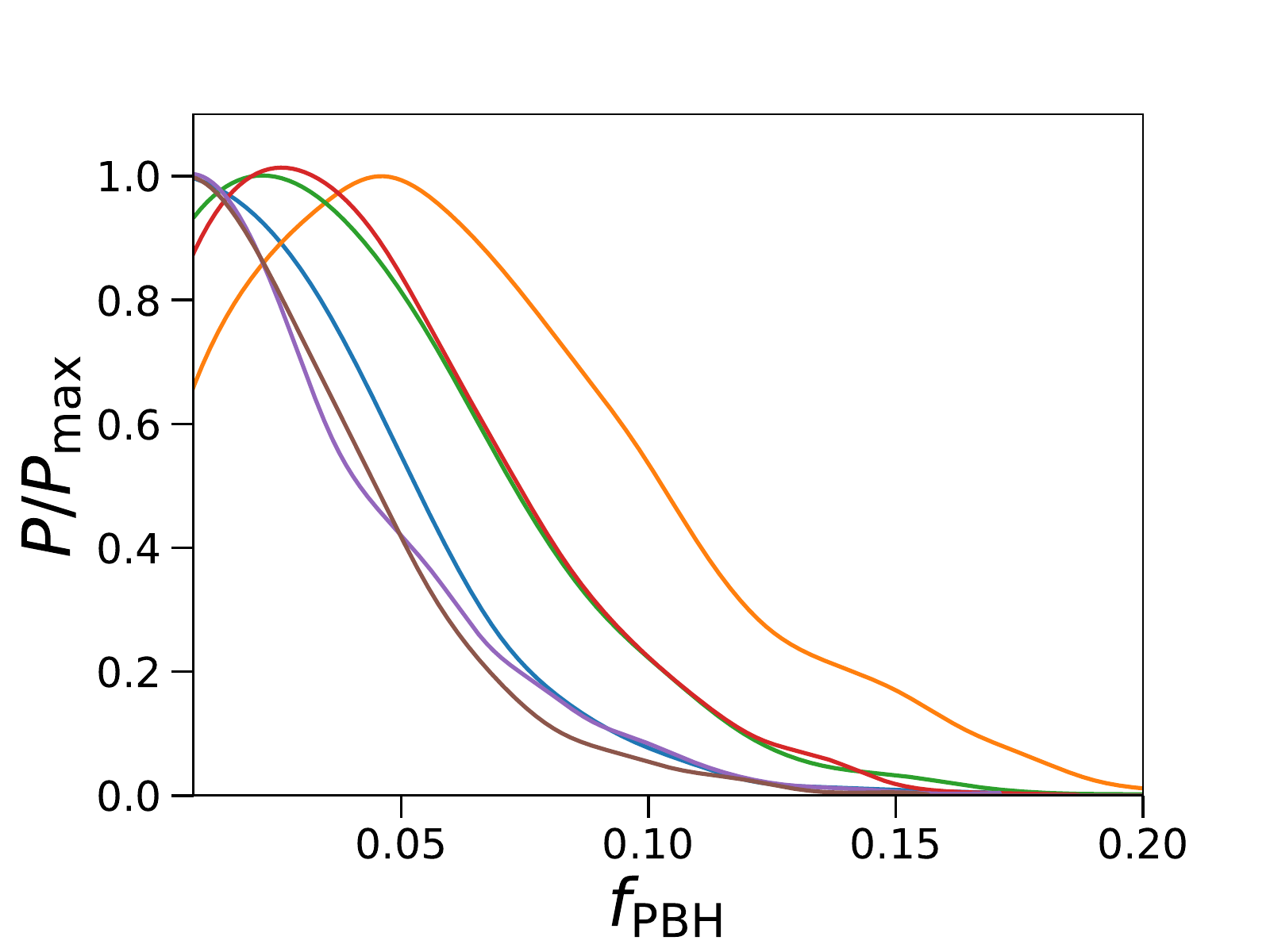}
\caption{Margnialized probability distribution of $\fpbh$ obtained assuming the extended cosmologies and those obtained assuming $\Lambda$CDM using PlanckTT+lowP+lensing for collisional ionization and 520$\Msun$ (left) and photoionization $100\Msun$ (right).}
\label{fig:f1d_ext}
\end{figure}

Fig. \ref{fig:f1d_ext} shows how the degeneracies in extended cosmologies weaken the constraints on $\fpbh$. Adding an extra cosmological parameter to the 6-$\Lambda$CDM ones, weakens the constraints on $\fpbh$ approximately by a factor 2, except for $w$CDM, due to the lack of correlation. 
The shift of the parameters $w$ and $\beta$ from their $\Lambda$CDM-fiducial values is responsible for making the $\fpbh$ constraints in the extended model slightly tighter.

\subsection{Constraints on PBHs with extended mass distributions}\label{sec:EMF}
So far, we have considered monochromatic populations of PBHs. Although this is an interesting first step, a more realistic case involves a distribution of masses. 
We follow \cite{bellomo:pbhemfconstraints} as described in Section \ref{sec:EMDmeth} and assess numerically the accuracy of the approximation in six specific cases: two Lognormal distributions and four Power Law distributions (two with $\gamma=0$ and other two with $\gamma=-0.5$). For each EMD case, we consider a narrow and a wide distribution. Our aim is to test the approach of \cite{bellomo:pbhemfconstraints} by comparing the results obtained following their prescription with an exact calculation using MCMCs.

\begin{figure}[h!]
\centering
\includegraphics[width=\linewidth]{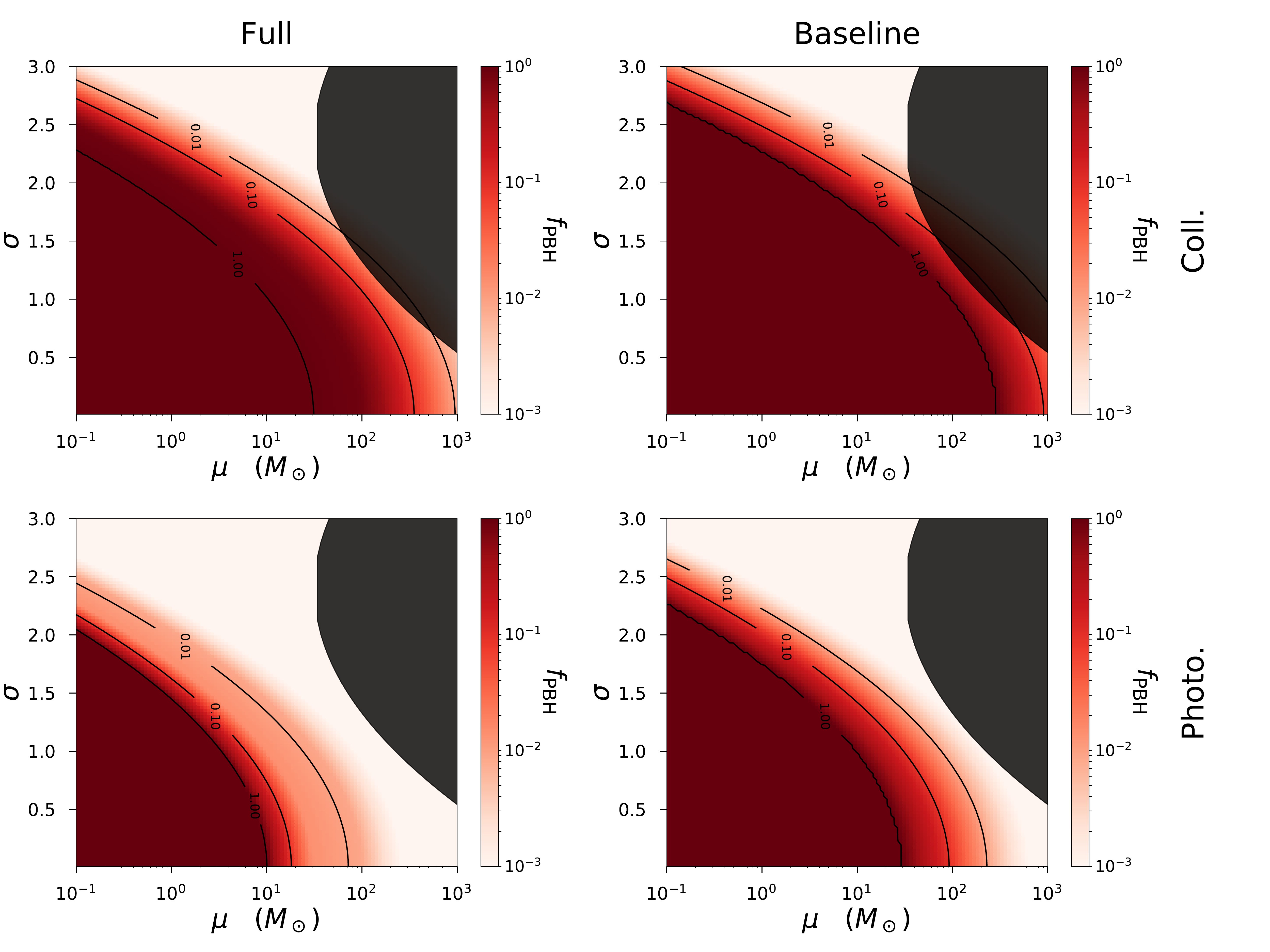}
\caption{95$\%$ confidence level marginalized constraints on $\fpbh$ (in color code) for a Lognormal mass distribution as function of the two parameters of the distribution, $\mu$ and $\sigma$, in the collisional regime (upper panels) and in the photoionization regime (bottom panels). The black shaded region indicates the values of $\mu$ and $\sigma$ for which EMDs extends beyond $10^4\Msun$ (the ratio between the values of the distribution in $M=10^4\Msun$ and in the peak is $10^{-5}$), masses for which AHK formalism breaks down and Eq.~\eqref{eq:cmb_equivalence} is not valid. Left panels: Results obtained using full Planck. Right panels: results obtained without including the high $\ell$ of polarization. }
\label{fig:2D_LN}
\end{figure}

\begin{figure}[h!]
\centering
\includegraphics[width=\linewidth]{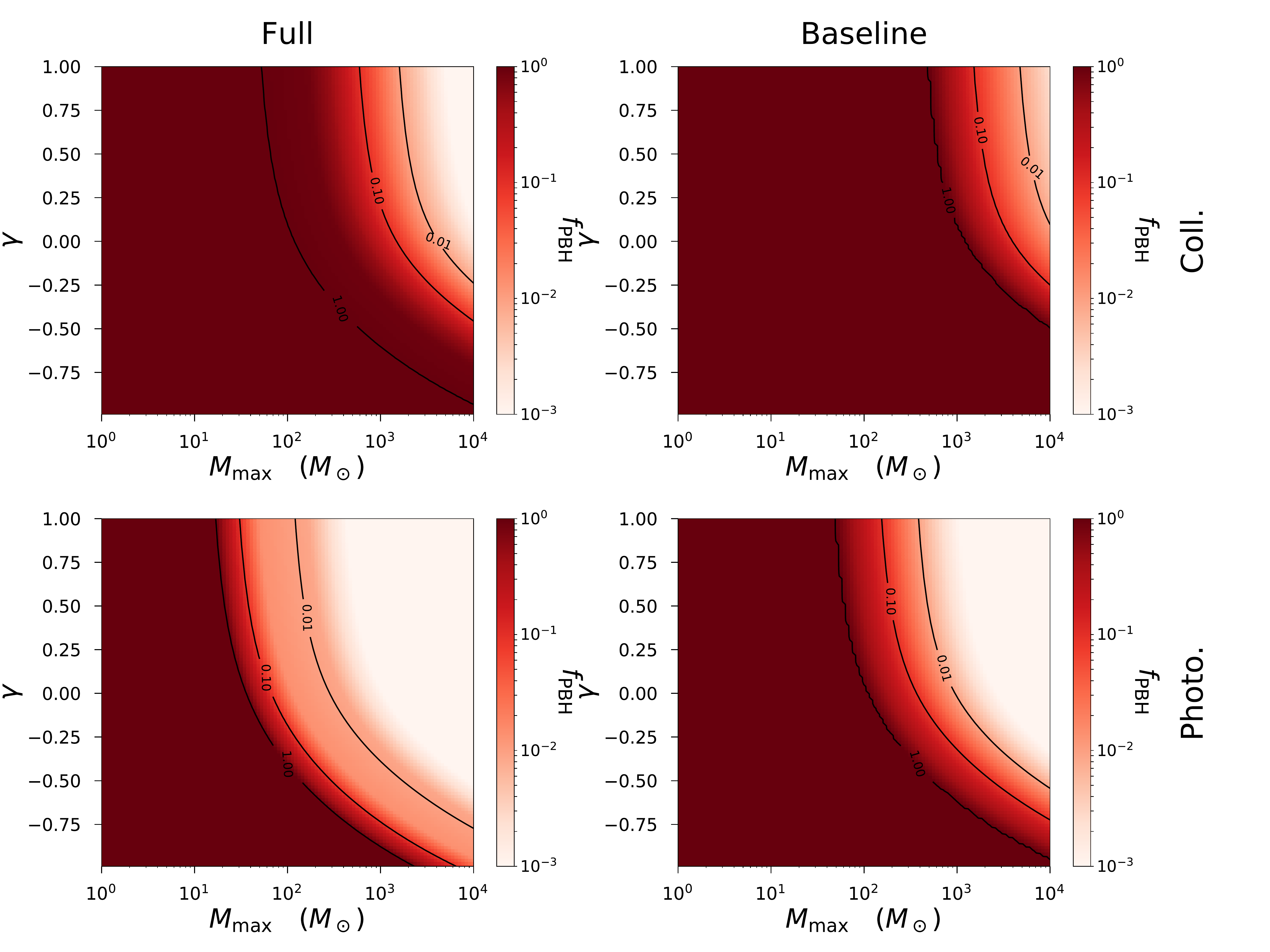}
\caption{95$\%$ confidence level marginalized constraints on $\fpbh$ (in color code) for a Power Law mass distribution as function of two parameters of the distribution, $M_{\rm max}$ and $\gamma$, for a fixed $M_{\rm min} = 10^{-2}\Msun$ in the collisional regime (upper panels) and in the photoionization regime (bottom panels). Left panels: Results obtained using full Planck. Right panels: results obtained without including the high $\ell$ of polarization. 
}
\label{fig:2D_PL}
\end{figure}

The resulting marginalized 68$\%$ and 95$\%$ confidence level upper limits on $\fpbh$ using full Planck for the six EMDs are shown in Table \ref{tab:EMF}. In the second column we report results for a monochromatic distribution of $30 \Msun$ and in the following columns we report constraints for extended mass distributions chosen to have an effective equivalent mass (according to Sec.~\ref{sec:EMDmeth}) of $30 \Msun$. The differences in the upper limits are well below the theoretical uncertainties of the model. These results demonstrates that, within the current knowledge of how an abundant population of PBHs affects the CMB, there is no difference between computing the constraints using the actual EMD and using an effective equivalent mass following the approach of \cite{bellomo:pbhemfconstraints}. 
\begin{table}[h]
\small
\begin{center}
\begin{tabular}{|c|c|c|c|c|c|c|c|}
\hline
C.L.	& M=30$\Msun$	& LN (1)	& LN (2)	& PL, (1) & PL, (2)	& PL, (1) & PL, (2) \\
  & $\equiv M_{\rm eq}$	& 	& 	& $\gamma = 0$ & $\gamma = 0$ 	& $\gamma = -0.5$ & $\gamma = -0.5$ \\
\hline
$\fpbh \leq (68\%)$	& 0.071	& 0.072	&	0.067	 & 0.073	&	0.070	& 0.069	& 0.0725	\\
\hline
$\fpbh \leq (95\%)$	& 0.14	& 0.15	&	0.14	& 0.14	& 0.14	&	0.14 & 0.16	\\
\hline
\end{tabular}
\end{center}
\caption{
Evaluation of the performance of the effective equivalent mass to match the results of accounting properly for  EMDs. The result using the  monochromatic  approximation is shown in the second column (M=30$\Msun$), results obtained using  several extended distributions are shown in the following columns. We compare the 68$\%$ and 95$\%$ marginalized upper limits on $\fpbh$ using full Planck and in the photoionization regime for six different EMDs with an equivalent mass of $30\Msun$ and the correspondent monochromatic case. We consider lognormal distributions (LN) with $\alpha = 0.2$ and $\{\mu,\sigma\} = \{22.8\Msun,0.5\}$ (1) and $\{\mu,\sigma\} = \{10\Msun,1.0\}$~(2). We also consider power law distributions with $\alpha= 0.3$ and $\gamma = 0$ with $\{M_{\rm min},M_{\rm max}\}=\{1\Msun,82\Msun\}$ (1) and $\{M_{\rm min},M_{\rm max}\}=\{0.01\Msun,114\Msun\}$ (2), and $\gamma = -0.5$ with $\{M_{\rm min},M_{\rm max}\}=\{1\Msun,150\Msun\}$ (1) and $\{M_{\rm min},M_{\rm max}\}=\{0.01\Msun,564\Msun\}$ (2). Note how similar the constraints are and how accurate is the monochromatic approximation.
}
\label{tab:EMF}
\end{table}
In the same spirit, the monochromatic upper limits on $\fpbh$ can be used to constrain the parameters of the EMDs considered here; i.e., $\mu$ and $\sigma$ for the Lognormal and $\gamma$ for the Power Law (for which we consider different mass bounds). We use the marginalized 95$\%$ confidence level upper limits of $\fpbh$ and translate these into limits for the EMDs parameters as follows. For every choice of EMD and its parameters, we compute the effective equivalent monochromatic mass and look up the $\fpbh$ value corresponding to the 95\% confidence upper limit (see Sec.\ref{sec:MethandData} and Fig.~\ref{fig:fpbh_constraints}). 
The results for photoionization and collisional ionization limits using both full Planck and PlanckTT+lowP+lensing for a $\Lambda$CDM model are shown in Figure \ref{fig:2D_LN} for a Lognormal EMD and Figure \ref{fig:2D_PL} for Power Law EMDs. The black shaded region indicates parameters values for which the EMD extends significantly beyond the limit of $10^4 \Msun$ where the AHK formalism becomes invalid. 

The above results can be understood as follows.
If PBHs have an EMD, the largest contribution to the injected energy comes from the high mass tail \cite{bellomo:pbhemfconstraints}. Therefore for a large width of a Lognormal mass distribution, $\mu$ is constrained to be small. In the case of the Power Law mass distribution, the flatter the distribution (i.e., the larger $\gamma$), the stronger the constraint on $\fpbh$. In essence, this implies that the constraints on $\fpbh$ for EMDs are stronger than for monochromatic distributions. Then, if PBHs are meant to be all the dark matter and have an EMD, this has to be peaked at $M\lesssim 1-50\Msun$, hence the allowed window shrinks.

From  Eq.~\eqref{eq:cmb_equivalence} and the results shown in Tab.~\ref{tab:fpbh_mono} it is possible to define a region in the  parameter space of the EMDs for which $\fpbh\sim 1$ is allowed (dark red in Figures \ref{fig:2D_LN} and \ref{fig:2D_PL}). Let us define $M_{\rm lim}$ to be  the maximum mass of a monochromatic distribution for which there are no upper limits on $\fpbh$ (values reported in Tab.~ \ref{tab:fpbh_mono}).  We can then  constrain the parameter space of EMDs by imposing $M_{\rm eq} \lesssim M_{\rm lim}$. For Lognormal distributions, there is an analytic solution:
\begin{equation}
\left\lbrace
\begin{aligned}
& \mu \lesssim M_{\rm lim},\\
& \sigma \lesssim \sqrt{\frac{2}{2+\alpha}\log \left(\frac{M_{\rm lim}}{\mu}\right)}.
\end{aligned}\right.
\end{equation}

\section{Discussion and conclusions}\label{sec:Conclusions}
The presence of a population of PBHs constituting a large fraction of the dark matter (i.e., in what we call $\Lambda$PBH model) would have injected radiation in the primordial plasma affecting the Universe's thermal and ionization histories and leaving an imprint on e.g. the CMB. A robustly demonstrated absence of these signatures would put strong limits on the abundance of PBHs of masses $\gtrsim 10\Msun$ and thus help to exclude a possible dark matter candidate. We find no evidence for the presence of such PBHs with an abundance large enough to be appreciable: $\fpbh=0$ is always inside the $68\%$ confidence region in all the cases considered. However, the sensitivity of present experiments is still not good enough to rule out PBHs in this mass range as a sizeable fraction of the dark matter.
Moreover, on the theoretical side, uncertainties in the modelling of processes such as accretion mechanisms of PBH or velocity distribution imply that any  constraint should be interpreted as an order of magnitude estimate, rather than a precise quantity. 

 In this work we use the formalism introduced in AHK~\cite{Ali-Haimoud_PBH}, which assumes spherical accretion and two limiting scenarios in which either collisional or photoionization processes completely dominate. We build on AHK by performing a robust statistical analysis, considering extended mass distributions of PBHs and exploring cosmological parameter degeneracies and the cosmological consequences that a large $\fpbh$ would have on other parameters.
Beside confirming AHK findings, we note that  the current allowed window of $\fpbh\sim 1$ for monochromatic populations of PBHs with masses $\sim 10-100\Msun$ greatly depends on the dataset considered (see Figure \ref{fig:fpbh_constraints} and Tab.~\ref{tab:fpbh_mono}). A significant component of the constraining power of CMB observations comes from the high multipoles of polarization power spectrum: the marginalized constraints on $\fpbh$ are about ten times stronger when included.

For Planck's recommended baseline data (i.e., not including the high multipoles of polarization power spectrum), CMB observations allow $\fpbh\sim 1$ for $M\lesssim 30 \Msun$ for the most stringent case (photoionization) and $M\lesssim 300 \Msun$ for the most conservative one (collisional ionization). The inclusion of high$\ell$ polarization data shrinks significantly this window to $M\lesssim 10 \Msun$ for photoionization and $M\lesssim 30 \Msun$ for collisional ionization.
The forthcoming release of Planck data with confirmation that the high$\ell$ polarization power spectrum is suitable to be used in extended cosmologies will greatly constrain the models for PBHs and their role as a dark matter component.

We have also investigated degeneracies between $\fpbh$ for  PBH masses in the  allowed $\Lambda$PBH model window of $\sim {\cal O}(10)\, \Msun$ to $\sim {\cal O}(500) \, \Msun$ and some cosmological parameters in a $\Lambda$PBH  cosmology (Figures \ref{fig:coll_degen} and \ref{fig:photo_degen}). 
As expected, we find a large correlation with $n_s$,  which gets stronger for larger $M_{\rm PBH}$, and allows even $n_s \approx 1$. Since future galaxy surveys will measure $n_s$ with astonishing precision, it will become possible to limit this degeneracy and set an indirect constraint on $\fpbh$. 
Interestingly, we find that, compared to the standard $\fpbh=0$ model, a sizeable $\fpbh$ would yield a higher sound horizon at radiation drag, $r_{\rm s}$.
A joint analysis with BAO would yield a lower value for $H_0$. Therefore a $\Lambda$PBH Universe do not ease some of the existing tensions of $\Lambda$CDM (and in particular the $H_0$ one, see e.g.,~\cite{BernalH0}), and possibly even worsen them.

However, a sizeable $\fpbh$
, if ignored, would bias the determination of $r_{\rm s}$ by $\sim 1$ Mpc. 
It is important to note that this shift is non-negligible compared to expected errors in the determination of the BAO distance scale, hence this possibility should be kept in mind when interpreting forthcoming BAO data. 

We have shown that the ratios between marginalized upper limits of $\fpbh$ for two different data sets do not depend on the assumptions about the ionization regime. Moreover, the ratios between the constraints on $\fpbh$ for different ionization regimes do not depend on the inclusion of the high multipoles of polarization power spectrum (Tab.~\ref{tab:rel_const_lcdm}).  
This suggests that using the ratio between marginalized upper limits on $\fpbh$ it is possible to isolate effects of specific choices of modelling or data sets, hence our findings can be reinterpreted for different modelling approaches, such as the disk accretion modelling of \cite{poulin:cmbconstraint}. 

\begin{figure}[h]
\centering
\includegraphics[width=0.45\textwidth]{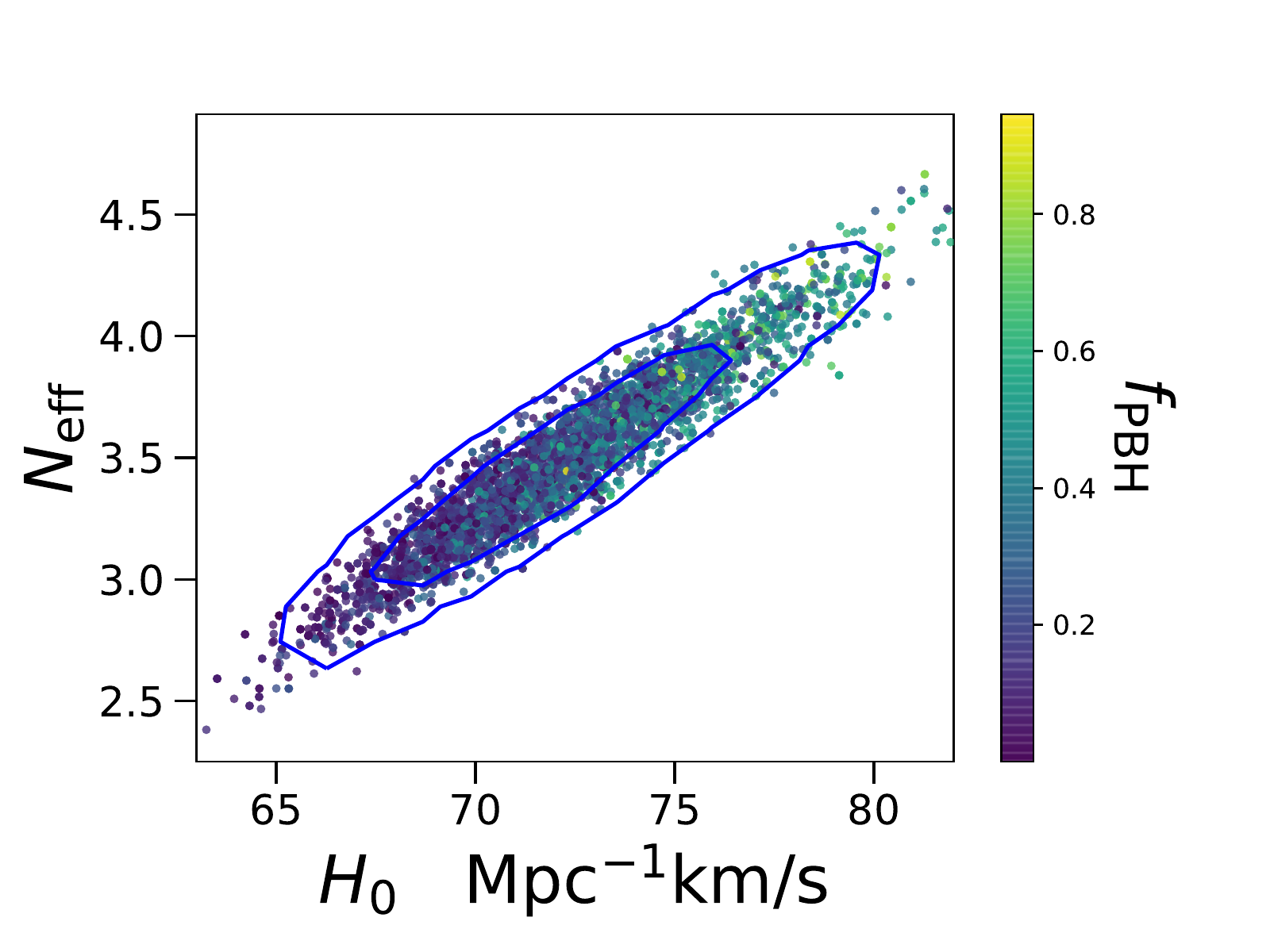}
\includegraphics[width=0.45\textwidth]{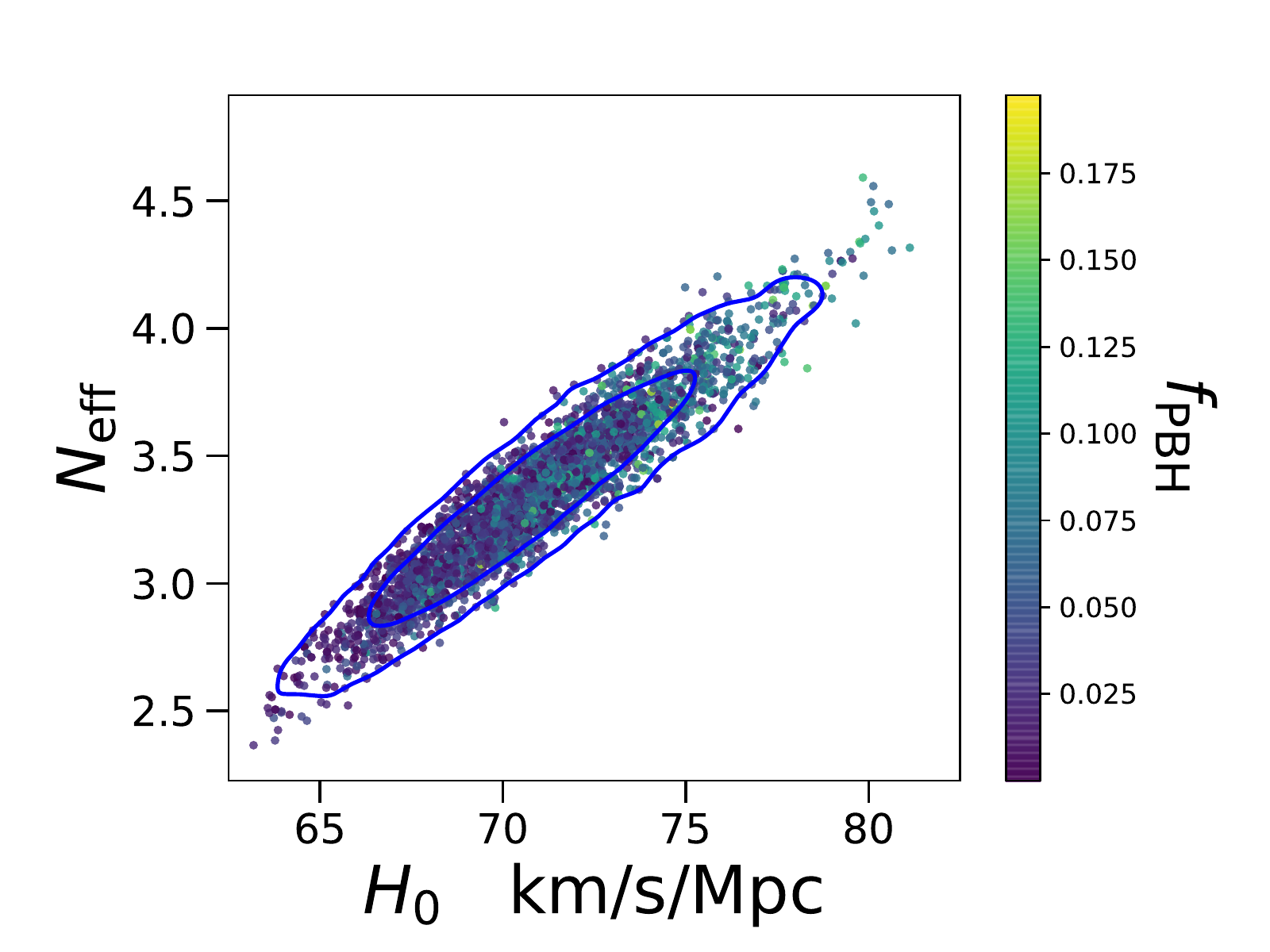}
\caption{68$\%$ and 95$\%$ constraints on the plane $N_{\rm eff}$-$H_0$ with a color code to express the value of $\fpbh$ for collisional ionization and 520$M_\odot$ (left) and for photoionization and 100$\Msun$ (right) using the baseline Planck data set. The marginalized constraints on the plane $N_{\rm eff}$-$\fpbh$ can be seen in figure \ref{fig:extcosmo}.}
\label{fig:NeffH0fpbh}
\end{figure}

If PBHs as dark matter are studied in the framework of popular extensions to $\Lambda$CDM, the allowed parameter space is extended because of degeneracies between $\fpbh$ and i.e., $N_{\rm eff}$ or the neutrino mass (Figure \ref{fig:extcosmo}). These degeneracies make it possible that PBH as dark matter might be hidden when assuming a $\Lambda$CDM background model, by forcing extended parameters to its fiducial values. The degeneracies with the parameters related with the neutrino sector can be explained by the fact that exotic neutrino physics would change the expansion history of the early Universe, therefore changing the epoch of recombination. A larger $\fpbh$ requires a larger $N_{\rm eff}$, which yields a higher $H_0$ CMB-inferred value when $\Lambda$CDM+$N_{\rm eff}$ is assumed, as can be seen in Figure \ref{fig:NeffH0fpbh}. This could make the CMB inferred value of $H_0$ fully compatible with direct measurements \cite{BennettH0_2014, EfstathiouH0_2014, BernalH0,Freedman_H0}. 

Finally, we successfully test the approach proposed in \cite{bellomo:pbhemfconstraints} to convert constraints computed for monochromatic distributions into constraints for EMDs in the case of CMB observations. Given the simplicity of the approach and the performance, we advocate using it to quickly get precise estimate for any EMD.  We also present constraints on the properties of popular EMDs, as to being consistent with current CMB data.

 Following this approach, we compute 95$\%$ confidence level upper limits on $\fpbh$ as function of the parameters of the EMDs (Figures \ref{fig:2D_LN} and \ref{fig:2D_PL}). 
 We find that CMB sets strong constraints on $\fpbh$ for EMDs that extends towards large masses. If one takes into account microlensing constraints (which constrain PBHs in the $M\lesssim 1-10 \Msun$ range), $\fpbh \sim 1$ and thus  a $\Lambda$PBH cosmology is allowed only for narrow EMDs and only if most conservative assumptions and data are considered.

We eagerly await the release of the final Planck high$\ell$ polarization data: in combination with other constraints it will be key to boost or rule out the hypothesis of PBHs as dark matter. 

\acknowledgments
We thank Yacine Ali-Ha\"imoud for sharing his modified version of HyRec with us and for valuable discussions which help to improve this work. We also thank Valerie Domcke, Tommi Tenkanen, Julian B. Mu\~ noz, Enrico Barausse and Ely Kovetz for useful comments on this manuscript. 
Funding for this work was partially provided by the Spanish MINECO under projects AYA2014-58747-P AEI/FEDER UE and MDM-2014-0369 of ICCUB (Unidad de Excelencia Maria de Maeztu).
JLB is supported by the Spanish MINECO under grant BES-2015-071307, co-funded by the ESF. NB is supported by the Spanish MINECO under grant BES-2015-073372. AR has received funding from the People Programme (Marie Curie Actions) of the European Union H2020 Programme under REA grant agreement number 706896 (COSMOFLAGS). LV acknowledges support of  European Union's Horizon 2020 research and innovation programme (BePreSySe, grant agreement 725327) and thanks the Radcliffe  Institute for Advanced Study at Harvard University for hospitality.

Based on observations obtained with Planck (http://www.esa.int/Planck), an ESA science mission with instruments and contributions directly funded by ESA Member States, NASA, and Canada.

\bibliography{biblio}

\providecommand{\href}[2]{#2}\begingroup\raggedright\begin{thebibliography}{10}

\bibitem{Zeldovich_pbh}
Y.~B. {Zel'dovich} and I.~D. {Novikov}, ``{The Hypothesis of Cores Retarded
  during Expansion and the Hot Cosmological Model},'' {\em Soviet Astronomy}
  {\bfseries 10} (Feb., 1967) 602.

\bibitem{Chapline_pbh}
G.~F. {Chapline}, ``{Cosmological effects of primordial black holes},''
  \href{http://dx.doi.org/10.1038/253251a0}{{\em Nature} {\bfseries 253} (Jan.,
  1975) 251}.

\bibitem{Alcock_micro1998}
C.~{Alcock}, R.~A. {Allsman}, and e.~a. {Alves}, ``{EROS and MACHO Combined
  Limits on Planetary-Mass Dark Matter in the Galactic Halo},''
  \href{http://dx.doi.org/10.1086/311355}{{\em Astrophys. J. Letter} {\bfseries
  499} (May, 1998) L9--L12},
  \href{http://arxiv.org/abs/astro-ph/9803082}{{\ttfamily astro-ph/9803082}}.

\bibitem{Flynn_barydm97}
C.~{Flynn}, A.~{Gould}, and J.~N. {Bahcall}, ``{Hubble Deep Field Constraint on
  Baryonic Dark Matter},'' \href{http://dx.doi.org/10.1086/310174}{{\em
  Astrophys. J. Letter} {\bfseries 466} (Aug., 1996) L55},
  \href{http://arxiv.org/abs/astro-ph/9603035}{{\ttfamily astro-ph/9603035}}.

\bibitem{Carr_dynamic1999}
B.~J. {Carr} and M.~{Sakellariadou}, ``{Dynamical Constraints on Dark Matter in
  Compact Objects},'' \href{http://dx.doi.org/10.1086/307071}{{\em Astrophys.
  J.} {\bfseries 516} (May, 1999) 195--220}.

\bibitem{Wilkinson_smbhconstraints}
P.~N. {Wilkinson}, D.~R. {Henstock}, I.~W. {Browne}, A.~G. {Polatidis},
  P.~{Augusto}, A.~C. {Readhead}, T.~J. {Pearson}, W.~{Xu}, G.~B. {Taylor}, and
  R.~C. {Vermeulen}, ``{Limits on the Cosmological Abundance of Supermassive
  Compact Objects from a Search for Multiple Imaging in Compact Radio
  Sources},'' \href{http://dx.doi.org/10.1103/PhysRevLett.86.584}{{\em Physical
  Review Letters} {\bfseries 86} (Jan., 2001) 584--587},
  \href{http://arxiv.org/abs/astro-ph/0101328}{{\ttfamily astro-ph/0101328}}.

\bibitem{Jungman_wimps}
G.~{Jungman}, M.~{Kamionkowski}, and K.~{Griest}, ``{Supersymmetric dark
  matter},'' \href{http://dx.doi.org/10.1016/0370-1573(95)00058-5}{{\em Phys.
  Rep.} {\bfseries 267} (Mar., 1996) 195--373},
  \href{http://arxiv.org/abs/hep-ph/9506380}{{\ttfamily hep-ph/9506380}}.

\bibitem{Arcadi_wimps}
G.~{Arcadi}, M.~{Dutra}, P.~{Ghosh}, M.~{Lindner}, Y.~{Mambrini}, M.~{Pierre},
  S.~{Profumo}, and F.~S. {Queiroz}, ``{The Waning of the WIMP? A Review of
  Models, Searches, and Constraints},'' {\em ArXiv e-prints} (Mar., 2017) ,
  \href{http://arxiv.org/abs/1703.07364}{{\ttfamily arXiv:1703.07364
  [hep-ph]}}.

\bibitem{abbott:ligo}
{\bfseries LIGO Scientific Collaboration and Virgo Collaboration}
  Collaboration, B.~P. Abbott {\em et~al.}, ``Observation of Gravitational
  Waves from a Binary Black Hole Merger,''
  \href{http://dx.doi.org/10.1103/PhysRevLett.116.061102}{{\em Phys. Rev.
  Lett.} {\bfseries 116} (Feb, 2016) 061102},
  \href{http://arxiv.org/abs/1602.03837}{{\ttfamily 1602.03837}}.

\bibitem{Bird_pbh}
S.~{Bird}, I.~{Cholis}, J.~B. {Mu{\~n}oz}, Y.~{Ali-Ha{\"\i}moud},
  M.~{Kamionkowski}, E.~D. {Kovetz}, A.~{Raccanelli}, and A.~G. {Riess}, ``{Did
  LIGO Detect Dark Matter?},''
  \href{http://dx.doi.org/10.1103/PhysRevLett.116.201301}{{\em Physical Review
  Letters} {\bfseries 116} no.~20, (May, 2016) 201301},
  \href{http://arxiv.org/abs/1603.00464}{{\ttfamily arXiv:1603.00464}}.

\bibitem{Sasaki_pbh}
M.~{Sasaki}, T.~{Suyama}, T.~{Tanaka}, and S.~{Yokoyama}, ``{Primordial Black
  Hole Scenario for the Gravitational-Wave Event GW150914},''
  \href{http://dx.doi.org/10.1103/PhysRevLett.117.061101}{{\em Physical Review
  Letters} {\bfseries 117} no.~6, (Aug., 2016) 061101},
  \href{http://arxiv.org/abs/1603.08338}{{\ttfamily arXiv:1603.08338}}.

\bibitem{Clesse_cluspbh}
S.~{Clesse} and J.~{Garc{\'{\i}}a-Bellido}, ``{The clustering of massive
  Primordial Black Holes as Dark Matter: Measuring their mass distribution with
  advanced LIGO},'' \href{http://dx.doi.org/10.1016/j.dark.2016.10.002}{{\em
  Physics of the Dark Universe} {\bfseries 15} (Mar., 2017) 142--147},
  \href{http://arxiv.org/abs/1603.05234}{{\ttfamily arXiv:1603.05234}}.

\bibitem{Kohri_smbh}
K.~{Kohri}, T.~{Nakama}, and T.~{Suyama}, ``{Testing scenarios of primordial
  black holes being the seeds of supermassive black holes by ultracompact
  minihalos and CMB {$\mu$} distortions},''
  \href{http://dx.doi.org/10.1103/PhysRevD.90.083514}{{\em Phys. Rev. D}
  {\bfseries 90} no.~8, (Oct., 2014) 083514},
  \href{http://arxiv.org/abs/1405.5999}{{\ttfamily arXiv:1405.5999}}.

\bibitem{Bernal_SMBH}
J.~L. Bernal, A.~Raccanelli, J.~Silk, E.~Kovetz, and L.~Verde, ``Primordial
  Black Holes as Seeds of Super Massive Black Holes,'' {\em in prep} (2017) .

\bibitem{silk_dwarf}
J.~{Silk}, ``{Feedback by Massive Black Holes in Gas-rich Dwarf Galaxies},''
  \href{http://dx.doi.org/10.3847/2041-8213/aa67da}{{\em Astrophys. J. Letters}
  {\bfseries 839} (Apr., 2017) L13},
  \href{http://arxiv.org/abs/1703.08553}{{\ttfamily arXiv:1703.08553}}.

\bibitem{raccanelli:cross}
A.~Raccanelli, E.~D. Kovetz, S.~Bird, I.~Cholis, and J.~B. Munoz,
  ``{Determining the progenitors of merging black-hole binaries},''
  \href{http://dx.doi.org/10.1103/PhysRevD.94.023516}{{\em Phys. Rev.}
  {\bfseries D94} no.~2, (2016) 023516},
\href{http://arxiv.org/abs/1605.01405}{{\ttfamily arXiv:1605.01405
  [astro-ph.CO]}}.

\bibitem{raccanelli:radio}
A.~Raccanelli, ``{Gravitational wave astronomy with radio galaxy surveys},''
  \href{http://dx.doi.org/10.1093/mnras/stx835}{{\em Mon. Not. Roy. Astron.
  Soc.} {\bfseries 469} no.~1, (2017) 656--670},
\href{http://arxiv.org/abs/1609.09377}{{\ttfamily arXiv:1609.09377
  [astro-ph.CO]}}.

\bibitem{raccanelli:quantumgravityconstraint}
A.~Raccanelli, F.~Vidotto, and L.~Verde, ``Effects of primordial black holes
  quantum gravity decay on galaxy clustering,''
  \href{http://arxiv.org/abs/1708.02588}{{\ttfamily 1708.02588}}.

\bibitem{cholis}
I.~Cholis, E.~D. Kovetz, Y.~Ali-Ha{\"\i}moud, S.~Bird, M.~Kamionkowski, J.~B.
  Mu{\~n}oz, and A.~Raccanelli, ``{Orbital eccentricities in primordial black
  hole binaries},'' \href{http://dx.doi.org/10.1103/PhysRevD.94.084013}{{\em
  Phys. Rev. D} {\bfseries D94} no.~8, (2016) 084013},
\href{http://arxiv.org/abs/1606.07437}{{\ttfamily arXiv:1606.07437
  [astro-ph.HE]}}.

\bibitem{kovetz}
E.~D. Kovetz, I.~Cholis, P.~C. Breysse, and M.~Kamionkowski, ``{Black hole mass
  function from gravitational wave measurements},''
  \href{http://dx.doi.org/10.1103/PhysRevD.95.103010}{{\em Phys. Rev. D}
  {\bfseries D95} no.~10, (2017) 103010},
\href{http://arxiv.org/abs/1611.01157}{{\ttfamily arXiv:1611.01157
  [astro-ph.CO]}}.

\bibitem{munoz}
J.~B. Mu{\~n}oz, E.~D. Kovetz, L.~Dai, and M.~Kamionkowski, ``{Lensing of Fast
  Radio Bursts as a Probe of Compact Dark Matter},''
  \href{http://dx.doi.org/10.1103/PhysRevLett.117.091301}{{\em Phys. Rev.
  Lett.} {\bfseries 117} no.~9, (2016) 091301},
\href{http://arxiv.org/abs/1605.00008}{{\ttfamily arXiv:1605.00008
  [astro-ph.CO]}}.

\bibitem{carr:comparison2}
B.~Carr, M.~Raidal, T.~Tenkanen, V.~Vaskonen, and H.~Veerm\"ae, ``Primordial
  black hole constraints for extended mass functions,''
  \href{http://dx.doi.org/10.1103/PhysRevD.96.023514}{{\em Phys. Rev. D}
  {\bfseries 96} (Jul, 2017) 023514}. \url{1705.05567}.

\bibitem{Kuhnel_EMF}
F.~{K{\"u}hnel} and K.~{Freese}, ``{Constraints on primordial black holes with
  extended mass functions},''
  \href{http://dx.doi.org/10.1103/PhysRevD.95.083508}{{\em Phys. Rev. D}
  {\bfseries 95} no.~8, (Apr., 2017) 083508},
  \href{http://arxiv.org/abs/1701.07223}{{\ttfamily arXiv:1701.07223}}.

\bibitem{Schutz_pta}
K.~{Schutz} and A.~{Liu}, ``{Pulsar timing can constrain primordial black holes
  in the LIGO mass window},''
  \href{http://dx.doi.org/10.1103/PhysRevD.95.023002}{{\em Phys. Rev. D}
  {\bfseries 95} no.~2, (Jan., 2017) 023002},
  \href{http://arxiv.org/abs/1610.04234}{{\ttfamily arXiv:1610.04234}}.

\bibitem{AliHaimoud_merger}
Y.~{Ali-Ha{\"i}moud}, E.~D. {Kovetz}, and M.~{Kamionkowski}, ``{The merger rate
  of primordial-black-hole binaries},'' {\em ArXiv e-prints} (Sept., 2017) ,
  \href{http://arxiv.org/abs/1709.06576}{{\ttfamily arXiv:1709.06576}}.

\bibitem{Kovetz_gwpbh}
E.~D. {Kovetz}, ``{Probing Primordial-Black-Hole Dark Matter with Gravitational
  Waves},'' \href{http://dx.doi.org/10.1103/PhysRevLett.119.131301}{{\em Phys.
  Rev. Lett.} {\bfseries 119} (Sep, 2017) 131301},
  \href{http://arxiv.org/abs/1705.09182}{{\ttfamily arXiv:1705.09182}}.

\bibitem{griest:keplerconstraint}
K.~Griest, A.~M. Cieplak, and M.~J. Lehner, ``Experimental Limits on Primordial
  Black Hole Dark Matter from the First 2 yr of Kepler Data,''
  \href{http://dx.doi.org/10.1088/0004-637X/786/2/158}{{\em The Astrophysical
  Journal} {\bfseries 786} no.~2, (2014) 158},
  \href{http://arxiv.org/abs/1307.5798}{{\ttfamily 1307.5798}}.

\bibitem{niikura:microlensingconstraint}
H.~Niikura, M.~Takada, N.~Yasuda, R.~H. Lupton, T.~Sumi, S.~More, A.~More,
  M.~Oguri, and M.~Chiba, ``Microlensing constraints on $10^{-10}M_\odot$-scale
  primordial black holes from high-cadence observation of M31 with Hyper
  Suprime-Cam,'' \href{http://arxiv.org/abs/1701.02151}{{\ttfamily
  1701.02151}}.

\bibitem{tisserand:microlensingconstraint}
{\bfseries EROS-2} Collaboration, P.~Tisserand {\em et~al.}, ``Limits on the
  Macho content of the Galactic Halo from the EROS-2 Survey of the Magellanic
  Clouds,'' \href{http://dx.doi.org/10.1051/0004-6361:20066017}{{\em A\&A}
  {\bfseries 469} no.~2, (2007) 387--404},
  \href{http://arxiv.org/abs/astro-ph/0607207}{{\ttfamily astro-ph/0607207}}.

\bibitem{calchinovati:microlensingconstraint}
S.~Calchi~Novati, S.~Mirzoyan, P.~Jetzer, and G.~Scarpetta, ``Microlensing
  towards the SMC: a new analysis of OGLE and EROS results,''
  \href{http://dx.doi.org/10.1093/mnras/stt1402}{{\em Monthly Notices of the
  Royal Astronomical Society} {\bfseries 435} no.~2, (2013) 1582--1597},
  \href{http://arxiv.org/abs/1308.4281}{{\ttfamily 1308.4281}}.

\bibitem{mediavilla:microlensingconstraint}
E.~Mediavilla, J.~A. Munoz, E.~Falco, V.~Motta, E.~Guerras, H.~Canovas,
  C.~Jean, A.~Oscoz, and A.~M. Mosquera, ``Microlensing-based Estimate of the
  Mass Fraction in Compact Objects in Lens Galaxies,''
  \href{http://dx.doi.org/10.1088/0004-637X/706/2/1451}{{\em The Astrophysical
  Journal} {\bfseries 706} no.~2, (2009) 1451},
  \href{http://arxiv.org/abs/0910.3645}{{\ttfamily 0910.3645}}.

\bibitem{quinn:widebinaryconstraint}
D.~P. Quinn, M.~I. Wilkinson, M.~J. Irwin, J.~Marshall, A.~Koch, and
  V.~Belokurov, ``On the reported death of the MACHO era,''
  \href{http://dx.doi.org/10.1111/j.1745-3933.2009.00652.x}{{\em Monthly
  Notices of the Royal Astronomical Society: Letters} {\bfseries 396} no.~1,
  (2009) L11--L15}, \href{http://arxiv.org/abs/0903.1644}{{\ttfamily
  0903.1644}}.

\bibitem{brandt:ufdgconstraint}
T.~D. Brandt, ``Constraints on MACHO Dark Matter from Compact Stellar Systems
  in Ultra-faint Dwarf Galaxies,''
  \href{http://dx.doi.org/10.3847/2041-8205/824/2/L31}{{\em The Astrophysical
  Journal Letters} {\bfseries 824} no.~2, (2016) L31},
  \href{http://arxiv.org/abs/1605.03665}{{\ttfamily 1605.03665}}.

\bibitem{gaggero}
D.~Gaggero, G.~Bertone, F.~Calore, R.~M.~T. Connors, M.~Lovell, S.~Markoff, and
  E.~Storm, ``{Searching for Primordial Black Holes in the radio and X-ray
  sky},'' \href{http://dx.doi.org/10.1103/PhysRevLett.118.241101}{{\em Phys.
  Rev. Lett.} {\bfseries 118} no.~24, (2017) 241101},
\href{http://arxiv.org/abs/1612.00457}{{\ttfamily arXiv:1612.00457
  [astro-ph.HE]}}.

\bibitem{giesen:annihilatingdm}
G.~Giesen, J.~Lesgourgues, B.~Audren, and Y.~Ali-Ha{\"\i}moud, ``CMB photons
  shedding light on dark matter,''
  \href{http://dx.doi.org/10.1088/1475-7516/2012/12/008}{{\em Journal of
  Cosmology and Astroparticle Physics} {\bfseries 2012} no.~12, (2012) 008},
  \href{http://arxiv.org/abs/https://arxiv.org/abs/1209.0247}{{\ttfamily
  https://arxiv.org/abs/1209.0247}}.

\bibitem{Ali-Haimoud_PBH}
Y.~{Ali-Ha{\"\i}moud} and M.~{Kamionkowski}, ``{Cosmic microwave background
  limits on accreting primordial black holes},''
  \href{http://dx.doi.org/10.1103/PhysRevD.95.043534}{{\em Phys. Rev. D}
  {\bfseries 95} no.~4, (Feb., 2017) 043534},
  \href{http://arxiv.org/abs/1612.05644}{{\ttfamily arXiv:1612.05644}}.

\bibitem{Planckparameterspaper}
{\bfseries Planck} Collaboration, P.~A.~R. Ade {\em et~al.}, ``{Planck 2015
  results. XIII. Cosmological parameters},''
  \href{http://dx.doi.org/10.1051/0004-6361/201525830}{{\em A\&A} {\bfseries
  594} (2016) A13}, \href{http://arxiv.org/abs/1502.01589}{{\ttfamily
  arXiv:1502.01589}}.

\bibitem{ricotti:pbhcosmologicaleffect}
M.~Ricotti, J.~P. Ostriker, and K.~J. Mack, ``Effect of Primordial Black Holes
  on the Cosmic Microwave Background and Cosmological Parameter Estimates,''
  \href{http://dx.doi.org/10.1086/587831}{{\em The Astrophysical Journal}
  {\bfseries 680} no.~2, (2008) 829},
  \href{http://arxiv.org/abs/0709.0524}{{\ttfamily 0709.0524}}.

\bibitem{poulin:cmbconstraint}
V.~Poulin, P.~D. Serpico, F.~Calore, S.~Clesse, and K.~Kohri, ``Squeezing
  spherical cows: CMB bounds on disk-accreting massive Primordial Black
  Holes,'' \href{http://arxiv.org/abs/1707.04206}{{\ttfamily 1707.04206}}.

\bibitem{carr:pbhfrominhomogeneities}
B.~J. Carr, ``The primordial black hole mass spectrum,''
  \href{http://dx.doi.org/10.1086/153853}{{\em Astrophysical Journal}
  {\bfseries 201} (1975) 1}.
  \url{http://adsabs.harvard.edu/abs/1975ApJ...201....1C}.

\bibitem{hawking:pbhformation}
S.~W. Hawking, ``Black holes from cosmic strings,''
  \href{http://dx.doi.org/10.1016/0370-2693(89)90206-2}{{\em Physics Letters B}
  {\bfseries 231} (1989) 237}.

\bibitem{hawking:pbhfrombubbles}
S.~W. Hawking, I.~G. Moss, and J.~M. Stewart, ``Bubble collisions in the very
  early universe,'' \href{http://dx.doi.org/10.1103/PhysRevD.26.2681}{{\em
  Phys. Rev. D} {\bfseries 26} (Nov, 1982) 2681--2693}.

\bibitem{Clesse_pbhhybrid}
S.~{Clesse} and J.~{Garc{\'{\i}}a-Bellido}, ``{Massive primordial black holes
  from hybrid inflation as dark matter and the seeds of galaxies},''
  \href{http://dx.doi.org/10.1103/PhysRevD.92.023524}{{\em Phys. Rev. D}
  {\bfseries 92} no.~2, (July, 2015) 023524},
  \href{http://arxiv.org/abs/1501.07565}{{\ttfamily arXiv:1501.07565}}.

\bibitem{Choptuik_criticcollapse}
M.~W. {Choptuik}, ``{Universality and scaling in gravitational collapse of a
  massless scalar field},''
  \href{http://dx.doi.org/10.1103/PhysRevLett.70.9}{{\em Physical Review
  Letters} {\bfseries 70} (Jan., 1993) 9--12}.

\bibitem{Niemeyer_criticcollapse}
J.~C. {Niemeyer} and K.~{Jedamzik}, ``{Near-Critical Gravitational Collapse and
  the Initial Mass Function of Primordial Black Holes},''
  \href{http://dx.doi.org/10.1103/PhysRevLett.80.5481}{{\em Physical Review
  Letters} {\bfseries 80} (June, 1998) 5481--5484},
  \href{http://arxiv.org/abs/astro-ph/9709072}{{\ttfamily astro-ph/9709072}}.

\bibitem{Musco_numcc05}
I.~{Musco}, J.~C. {Miller}, and L.~{Rezzolla}, ``{Computations of primordial
  black-hole formation},''
  \href{http://dx.doi.org/10.1088/0264-9381/22/7/013}{{\em Classical and
  Quantum Gravity} {\bfseries 22} (Apr., 2005) 1405--1424},
  \href{http://arxiv.org/abs/gr-qc/0412063}{{\ttfamily gr-qc/0412063}}.

\bibitem{Musco_numcc09}
I.~{Musco}, J.~C. {Miller}, and A.~G. {Polnarev}, ``{Primordial black hole
  formation in the radiative era: investigation of the critical nature of the
  collapse},'' \href{http://dx.doi.org/10.1088/0264-9381/26/23/235001}{{\em
  Classical and Quantum Gravity} {\bfseries 26} no.~23, (Dec., 2009) 235001},
  \href{http://arxiv.org/abs/0811.1452}{{\ttfamily arXiv:0811.1452 [gr-qc]}}.

\bibitem{Musco_numcc13}
I.~{Musco} and J.~C. {Miller}, ``{Primordial black hole formation in the early
  universe: critical behaviour and self-similarity},''
  \href{http://dx.doi.org/10.1088/0264-9381/30/14/145009}{{\em Classical and
  Quantum Gravity} {\bfseries 30} no.~14, (July, 2013) 145009},
  \href{http://arxiv.org/abs/1201.2379}{{\ttfamily arXiv:1201.2379 [gr-qc]}}.

\bibitem{carr:comparison1}
B.~Carr, F.~K\"uhnel, and M.~Sandstad, ``Primordial black holes as dark
  matter,'' \href{http://dx.doi.org/10.1103/PhysRevD.94.083504}{{\em Phys. Rev.
  D} {\bfseries 94} (Oct, 2016) 083504},
  \href{http://arxiv.org/abs/1607.06077}{{\ttfamily 1607.06077}}.

\bibitem{bellomo:pbhemfconstraints}
N.~{Bellomo}, J.~L. {Bernal}, A.~{Raccanelli}, and L.~{Verde}, ``{Primordial
  Black Holes as Dark Matter: Converting Constraints from Monochromatic to
  Extended Mass Distributions},'' {\em ArXiv e-prints} (Sept., 2017) ,
  \href{http://arxiv.org/abs/1709.07467}{{\ttfamily arXiv:1709.07467}}.

\bibitem{green:lognormal}
A.~M. Green, ``Microlensing and dynamical constraints on primordial black hole
  dark matter with an extended mass function,''
  \href{http://dx.doi.org/10.1103/PhysRevD.94.063530}{{\em Phys. Rev. D}
  {\bfseries 94} (Sep, 2016) 063530},
  \href{http://arxiv.org/abs/https://arxiv.org/abs/1609.01143}{{\ttfamily
  https://arxiv.org/abs/1609.01143}}.

\bibitem{Lesgourgues:2011re}
J.~Lesgourgues, ``{The Cosmic Linear Anisotropy Solving System (CLASS) I:
  Overview},''
\href{http://arxiv.org/abs/1104.2932}{{\ttfamily arXiv:1104.2932
  [astro-ph.IM]}}.

\bibitem{Blas:2011rf}
D.~Blas, J.~Lesgourgues, and T.~Tram, ``{The Cosmic Linear Anisotropy Solving
  System (CLASS) II: Approximation schemes},''
  \href{http://dx.doi.org/10.1088/1475-7516/2011/07/034}{{\em JCAP} {\bfseries
  1107} (2011) 034},
\href{http://arxiv.org/abs/1104.2933}{{\ttfamily arXiv:1104.2933
  [astro-ph.CO]}}.

\bibitem{alihamoud:hyrec1}
Y.~Ali-Ha\"{\i}moud and C.~M. Hirata, ``Ultrafast effective multilevel atom
  method for primordial hydrogen recombination,''
  \href{http://dx.doi.org/10.1103/PhysRevD.82.063521}{{\em Phys. Rev. D}
  {\bfseries 82} (Sep, 2010) 063521},
  \href{http://arxiv.org/abs/https://arxiv.org/abs/1006.1355}{{\ttfamily
  https://arxiv.org/abs/1006.1355}}.

\bibitem{alihamoud:hyrec2}
Y.~Ali-Ha\"{\i}moud and C.~M. Hirata, ``HyRec: A fast and highly accurate
  primordial hydrogen and helium recombination code,''
  \href{http://dx.doi.org/10.1103/PhysRevD.83.043513}{{\em Phys. Rev. D}
  {\bfseries 83} (Feb, 2011) 043513},
  \href{http://arxiv.org/abs/https://arxiv.org/abs/1011.3758}{{\ttfamily
  https://arxiv.org/abs/1011.3758}}.

\bibitem{Audren13_mp}
B.~{Audren}, J.~{Lesgourgues}, K.~{Benabed}, and S.~{Prunet}, ``{Conservative
  constraints on early cosmology with MONTE PYTHON},''
  \href{http://dx.doi.org/10.1088/1475-7516/2013/02/001}{{\em JCAP} {\bfseries
  2} (Feb., 2013) 1}, \href{http://arxiv.org/abs/1210.7183}{{\ttfamily
  arXiv:1210.7183}}.

\bibitem{Planck15_likelihood}
{\bfseries Planck} Collaboration, N.~Aghanim {\em et~al.}, ``{Planck 2015
  results. XI. CMB power spectra, likelihoods, and robustness of parameters},''
  \href{http://dx.doi.org/10.1051/0004-6361/201526926}{{\em A\&A} {\bfseries
  594} (2016) A11}, \href{http://arxiv.org/abs/1507.02704}{{\ttfamily
  arXiv:1507.02704}}.

\bibitem{Planck_newHFI}
{\bfseries Planck} Collaboration, N.~Aghanim {\em et~al.}, ``Planck
  intermediate results - XLVI. Reduction of large-scale systematic effects in
  HFI polarization maps and estimation of the reionization optical depth,''
  \href{http://dx.doi.org/10.1051/0004-6361/201628890}{{\em A\&A} {\bfseries
  596} (2016) A107}, \href{http://arxiv.org/abs/1605.02985}{{\ttfamily
  arXiv:1605.02985}}.

\bibitem{Ricotti_accretion}
M.~{Ricotti}, ``{Bondi Accretion in the Early Universe},''
  \href{http://dx.doi.org/10.1086/516562}{{\em Astrophys. J.} {\bfseries 662}
  (June, 2007) 53--61}, \href{http://arxiv.org/abs/0706.0864}{{\ttfamily
  arXiv:0706.0864}}.

\bibitem{marti:gwpbhmergers}
M.~Raidal, V.~Vaskonen, and H.~Veerm\"{a}e, ``Gravitational Waves from
  Primordial Black Hole Mergers,''
  \href{http://dx.doi.org/10.1088/1475-7516/2017/09/037}{{\em Journal of
  Cosmology and Astroparticle Physics} {\bfseries 2017} no.~09, (2017) 037},
  \href{http://arxiv.org/abs/1707.01480}{{\ttfamily arXiv:1707.01480}}.

\bibitem{RiessH0_2016}
A.~G. {Riess}, L.~M. {Macri}, S.~L. {Hoffmann}, D.~{Scolnic}, S.~{Casertano},
  A.~V. {Filippenko}, B.~E. {Tucker}, M.~J. {Reid}, D.~O. {Jones}, J.~M.
  {Silverman}, R.~{Chornock}, P.~{Challis}, W.~{Yuan}, P.~J. {Brown}, and R.~J.
  {Foley}, ``{A 2.4\% Determination of the Local Value of the Hubble
  Constant},'' \href{http://dx.doi.org/10.3847/0004-637X/826/1/56}{{\em ApJ}
  {\bfseries 826} (July, 2016) 56},
  \href{http://arxiv.org/abs/1604.01424}{{\ttfamily arXiv:1604.01424}}.

\bibitem{BernalH0}
J.~L. {Bernal}, L.~{Verde}, and A.~G. {Riess}, ``{The trouble with H$_{0}$},''
  \href{http://dx.doi.org/10.1088/1475-7516/2016/10/019}{{\em JCAP} {\bfseries
  10} (Oct., 2016) 019}, \href{http://arxiv.org/abs/1607.05617}{{\ttfamily
  arXiv:1607.05617}}.

\bibitem{StandardQuantities}
L.~{Verde}, J.~L. {Bernal}, A.~F. {Heavens}, and R.~{Jimenez}, ``{The length of
  the low-redshift standard ruler},''
  \href{http://dx.doi.org/10.1093/mnras/stx116}{{\em MNRAS} {\bfseries 467}
  (May, 2017) 731--736}, \href{http://arxiv.org/abs/1607.05297}{{\ttfamily
  arXiv:1607.05297}}.

\bibitem{BennettH0_2014}
C.~L. {Bennett}, D.~{Larson}, J.~L. {Weiland}, and G.~{Hinshaw}, ``{The 1\%
  Concordance Hubble Constant},''
  \href{http://dx.doi.org/10.1088/0004-637X/794/2/135}{{\em Astrophys. J.}
  {\bfseries 794} (Oct., 2014) 135},
  \href{http://arxiv.org/abs/1406.1718}{{\ttfamily arXiv:1406.1718}}.

\bibitem{EfstathiouH0_2014}
G.~{Efstathiou}, ``{H$_{0}$ revisited},''
  \href{http://dx.doi.org/10.1093/mnras/stu278}{{\em Mon. Not. Roy. Astron.
  Soc.} {\bfseries 440} (May, 2014) 1138--1152},
  \href{http://arxiv.org/abs/1311.3461}{{\ttfamily arXiv:1311.3461}}.

\bibitem{Freedman_H0}
W.~L. {Freedman}, ``{Cosmology at a crossroads},''
  \href{http://dx.doi.org/10.1038/s41550-017-0121}{{\em Nature Astronomy}
  {\bfseries 1} (May, 2017) 0121}.

\end{thebibliography}\endgroup
\bibliographystyle{utcaps}

\end{document}